\pdfoutput=1
\documentclass[11pt,twoside,a4paper,cmspaper,final,collab]{cms-tdr}
\def\svnVersion{5bedbfd-D}\def\svnDate{2021/03/07}\def\cmsCernNoTag{CERN-EP-2020-110}\def\cmsCernDate{\today}\def\cmsMessage{"Published in Physics Letters B as \href{http://dx.doi.org/10.1016/j.physletb.2021.136188}{\doi{10.1016/j.physletb.2021.136188}.}"}
\begin{document}\cmsNoteHeader{BPH-20-001}

\newcommand{\phis}{\ensuremath{\phi_{\mathrm{s}}}\xspace}
\newcommand{\DGs}{\ensuremath{\Delta\Gamma_{\mathrm{s}}}\xspace}
\newcommand{\Gs}{\ensuremath{\Gamma_{\mathrm{s}}}\xspace}
\newcommand{\Dms}{\ensuremath{\Delta m_{\mathrm{s}}}\xspace}
\newcommand{\betas}{\ensuremath{\beta_\mathrm{s}}\xspace}
\newcommand{\ips}{{\ensuremath{\unit{ps}^{-1}}}}
\newcommand{\mrad}{\ensuremath{\unit{mrad}}}
\newcommand{\BsJpsiphi}{\ensuremath{\PBzs \to \PJGy\,\PGf }\xspace}
\newcommand{\BsJpsiphialt}{\ensuremath{\PBzs \to \PJGy\,\PGf(1020) }\xspace}
\newcommand{\mumu}{\ensuremath{\PGmp\PGmm}\xspace}
\newcommand{\KK}{\ensuremath{\PKp\PKm}\xspace}
\newcommand{\BJpsiK}{\ensuremath{\PBpm\to\PJGy\,\PKpm}\xspace}
\newcommand{\BJpsiKMM}{\ensuremath{\PBpm\to\PJGy\,\PKpm\to\mumu\,\PKpm}\xspace}
\newcommand{\BsJpsiKK}{\ensuremath{\PBzs\to\PJGy\,\KK}\xspace}
\newcommand{\BsJpsiphiMMKK}{\ensuremath{\PBzs\to\PJGy\,\PGf\to\mumu\,\KK}\xspace}
\newcommand{\BsJpsiphiMMKKalt}{\ensuremath{\PBzs\to\PJGy\,\PGf(1020) \to \mumu\,\KK}\xspace}
\newcommand{\theT}{\ensuremath{\theta_\mathrm{T}}\xspace}
\newcommand{\phiT}{\ensuremath{\varphi_\mathrm{T}}\xspace}
\newcommand{\psiT}{\ensuremath{\psi_\mathrm{T}}\xspace}
\newcommand{\dif}[2]{\mathrm{d}^{#1}#2}
\newcommand{\Azero}{\ensuremath{A_0}}
\newcommand{\Aperp}{\ensuremath{A_{\perp}}}
\newcommand{\Apara}{\ensuremath{A_{\parallel}}}
\newcommand{\Aswav}{\ensuremath{A_{\mathrm{S}}}}
\newcommand{\dzero}{\ensuremath{\delta_0}}
\newcommand{\dperp}{\ensuremath{\delta_{\perp}}}
\newcommand{\dpara}{\ensuremath{\delta_{\parallel}}}
\newcommand{\dswav}{\ensuremath{\delta_{\mathrm{S}}}}
\newcommand{\dswpd}{\ensuremath{\delta_{\mathrm{S}\perp}}}
\newcommand{\sqabs}[1]{\abs{#1}^2}
\newcommand{\sqsin}[1]{\sin^2{#1}}
\newcommand{\sqcos}[1]{\cos^2{#1}}
\newcommand{\tenpow}[1]{10^{#1}}
\newcommand{\Nev}{\ensuremath{48\,500}\xspace}
\newcommand{\Nevtot}{\ensuremath{65\,500}\xspace}
\newcommand{\intL}{\ensuremath{96.4\fbinv}\xspace}
\newcommand{\hps}{\ensuremath{\hslash\unit{ps}^{-1}}}
\newcommand{\Bsh}{\ensuremath{\PB^{\mathrm{H}}_{\mathrm{s}}}\xspace}
\newcommand{\Bsl}{\ensuremath{\PB^{\mathrm{L}}_{\mathrm{s}}}\xspace}
\newcommand{\CP}{\ensuremath{CP}\xspace}
\newcommand{\etag}{\ensuremath{\varepsilon_{\text{tag}}}\xspace}
\newcommand{\wtag}{\ensuremath{\omega_{\text{tag}}}\xspace}
\newcommand{\Ptag}{\ensuremath{P_{\text{tag}}}\xspace}
\newcommand{\D}{\ensuremath{{\mathcal{D}}}\xspace}
\newcommand{\wevt}{\ensuremath{\omega_{\text{evt}}}\xspace}
\newcommand{\wmeas}{\ensuremath{\omega_{\text{meas}}}\xspace}
\newcommand{\ct}{\ensuremath{ct}\xspace}
\newcommand{\sct}{\ensuremath{\sigma_{\ct}}\xspace}
\newcommand{\like}{\ensuremath{{\mathcal{L}}}\xspace}
\newcommand{\kstar}{\ensuremath{\PKstP{892}^0}\xspace}
\newcommand{\ksp}{\ensuremath{k_\mathrm{SP}}\xspace}
\newcommand{\phisSM}{\ensuremath{-36.96\,^{+0.72}_{-0.84}}\xspace}
\newcommand{\phisComb}{\ensuremath{-21 \pm 44\stat \pm 10 \syst\mrad}\xspace}
\newcommand{\DGsComb}{\ensuremath{0.1032 \pm 0.0095\stat \pm 0.0048\syst\ips}\xspace}
\hyphenation{pseu-do-ex-per-i-ment}
\providecommand{\cmsTable}[1]{\resizebox{\textwidth}{!}{#1}}
\newlength\cmsTabSkip\setlength{\cmsTabSkip}{1ex}
\ifthenelse{\boolean{cms@external}}{\providecommand{\cmsLeft}{upper\xspace}}{\providecommand{\cmsLeft}{left\xspace}}
\ifthenelse{\boolean{cms@external}}{\providecommand{\cmsRight}{lower\xspace}}{\providecommand{\cmsRight}{right\xspace}}
\ifthenelse{\boolean{cms@external}}{\providecommand{\suppMaterial}{the supplemental material [URL will be inserted by publisher]}}{\providecommand{\suppMaterial}{Appendix~\ref{sec:appendix}}}

\cmsNoteHeader{BPH-20-001}
\title{Measurement of the \CP-violating phase \texorpdfstring{\phis}{phis} in the \texorpdfstring{$\BsJpsiphiMMKKalt$}{Bs -> J/psi phi} channel in proton-proton collisions at $\sqrt{s} = 13\TeV$}

\date{\today}

\abstract{
The \CP-violating weak phase \phis and the decay width difference \DGs between the light and heavy \PBzs mass eigenstates are measured with the CMS detector at the LHC in a sample of \Nev\ reconstructed \BsJpsiphiMMKKalt events.
The measurement is based on a data sample corresponding to an integrated luminosity of \intL, collected in proton-proton collisions at $\sqrt{s} = 13\TeV$ in 2017--2018.
To extract the values of \phis and \DGs, a time-dependent and flavor-tagged angular analysis of the $\mumu\KK$ final state is performed.
The analysis employs a dedicated tagging trigger and a novel opposite-side muon flavor tagger based on machine learning techniques.
The measurement yields $\phis = -11 \pm 50\stat \pm 10\syst\mrad$ and $\DGs = 0.114 \pm 0.014\stat \pm 0.007\syst\ips$, in agreement with the standard model predictions.
When combined with the previous CMS measurement at $\sqrt{s} = 8\TeV$, the following values are obtained: $\phis = \phisComb$, $\DGs = \DGsComb$, a significant improvement over the 8\TeV result.}

\hypersetup{
pdfauthor={CMS Collaboration},
pdftitle={Measurement of the CP-violating phase phis in the Bs to J/psi phi(1020) to mu+mu-K+K- channel in proton-proton collisions at sqrt(s) = 13 TeV},
pdfsubject={CMS},
pdfkeywords={CMS, physics, oscillations, B physics}}

\maketitle

\section{Introduction}
\label{sec:introd}

Precision tests of the standard model (SM) of particle physics have become increasingly important, since no direct evidence for new physics has been found so far at the CERN LHC.
Decays of \PBzs mesons present important opportunities to probe the consistency of the SM.
In this Letter, a new measurement of the \CP-violating weak phase \phis and the decay width difference \DGs between the light (\Bsl) and heavy (\Bsh) \PBzs meson mass eigenstates is presented.
Charge-conjugate states are implied throughout, unless stated otherwise.

The weak phase \phis arises from the interference between direct \PBzs meson decays to a \CP eigenstate of $\PQc\PAQc\PQs\PAQs$ and decays through mixing to the same final state.
In the SM, \phis is related to the elements of the Cabibbo--Kobayashi--Maskawa matrix via $\phis \simeq -2 \beta_\mathrm{s} = -2\arg(-V_{\PQt\PQs} V_{\PQt\PQb}^{*} / V_{\PQc\PQs} V_{\PQc\PQb}^{*})$, neglecting penguin diagram contributions, where $\beta_\mathrm{s}$ is one of the angles of the unitary triangles.
The current best determination of $-2\betas$ comes from a global fit to experimental data on \PQb hadron and kaon decays. 
Assuming no physics beyond the SM (BSM) in the \PBzs mixing and decays, a $-2\betas$ value of $\phisSM \mrad$ is determined~\cite{ref:Charles11prd}.
New physics can modify this phase via the contribution of BSM particles to \PBzs mixing~\cite{Chiang:2009ev,ref:Artuso2015swg}.
Since the numerical value of \phis in the SM is known very precisely, even a small deviation from this value would constitute evidence of BSM physics.
The decay width difference between the \Bsl and \Bsh eigenstates, on the other hand, is predicted less precisely at $\DGs = 0.091 \pm 0.013\ips$~\cite{ref:Lenz2019lvd}.
Its measurement provides an important test for theoretical predictions and can be used to further constrain new-physics effects~\cite{ref:Lenz2019lvd}.

The weak phase \phis was first measured by the Fermilab Tevatron experiments~\cite{ref:d0Abazov08prl, ref:d0Abazov12prd, ref:cdfAaltonen08prl, ref:cdfAaltonen12prd, ref:cdfAaltonen12prl}, and then at the LHC by the ATLAS, CMS, and LHCb experiments~\cite{ref:atlasAad12jhep, ref:atlasAad14prd, ref:atlasAad16jhep, ref:Aad:2020jfw, ref:cmsKhachatryan16plb, ref:lhcbAaij12plb, ref:lhcbAaij13prd, ref:lhcbAaij14plb, ref:lhcbAaij15prl, ref:Aaij2019vot}, using \BsJpsiphialt (referred to as \BsJpsiphi in what follows), $\PBzs \to \JPsi\, \PfDzP{980}$, and $\PBzs \to \JPsi\,\PSh^+ \PSh^-$ decays, where $\PSh$ stands for a kaon or pion.
Measurements of \phis in \PBzs decays to $\PGyP{2S}\Pgf$ and $\PDps\PDms$ were performed by the LHCb Collaboration~\cite{ref:lhcbAaij16plb, ref:lhcbAaij14prl}.

In this Letter, CMS results on the \BsJpsiphi decay to the $\mumu\KK$ final state are presented, and possible additional contributions to this final state from the $\PBzs \to\PJGy\, \PfDzP{980}$ and nonresonant \BsJpsiKK decays are taken into account by including a term for an additional $S$-wave amplitude in the decay model.
Compared to our previous measurement~\cite{ref:cmsKhachatryan16plb} at $\sqrt{s} =8\TeV$, we benefit from the increase in the center-of-mass energy from 8 to 13\TeV that nearly doubles the \PBzs production cross section and a novel opposite-side (OS) muon flavor tagger.
The new tagger employs machine learning techniques and achieves better discrimination power than previous methods.
We also make use of a specialized trigger that requires an additional (third) muon, which can be used for flavor tagging, improving the tagging efficiency at the cost of a reduced number of signal events.
As a result, the new measurement, while based on a similar number of \PBzs candidates as the earlier one~\cite{ref:cmsKhachatryan16plb}, allows us to double the precision in the determination of \phis, as well as measure some of the parameters that were constrained to their world-average values in our previous work~\cite{ref:cmsKhachatryan16plb}.
At the same time, the precision on parameters that do not benefit from the tagging information, such as \DGs, is comparable to that in the previous measurement.

Final states that are mixtures of \CP eigenstates require an angular analysis to separate the \CP-odd and \CP-even components.
A time-dependent angular analysis can be performed by measuring the decay angles of the final-state particles and the proper decay length of the reconstructed \PBzs candidate, which is equal to the proper decay time $t$ multiplied by the speed of light, and referred to as \ct in what follows.

\begin{figure}[hbtp]
  \centering
  \includegraphics[width=0.48\textwidth]{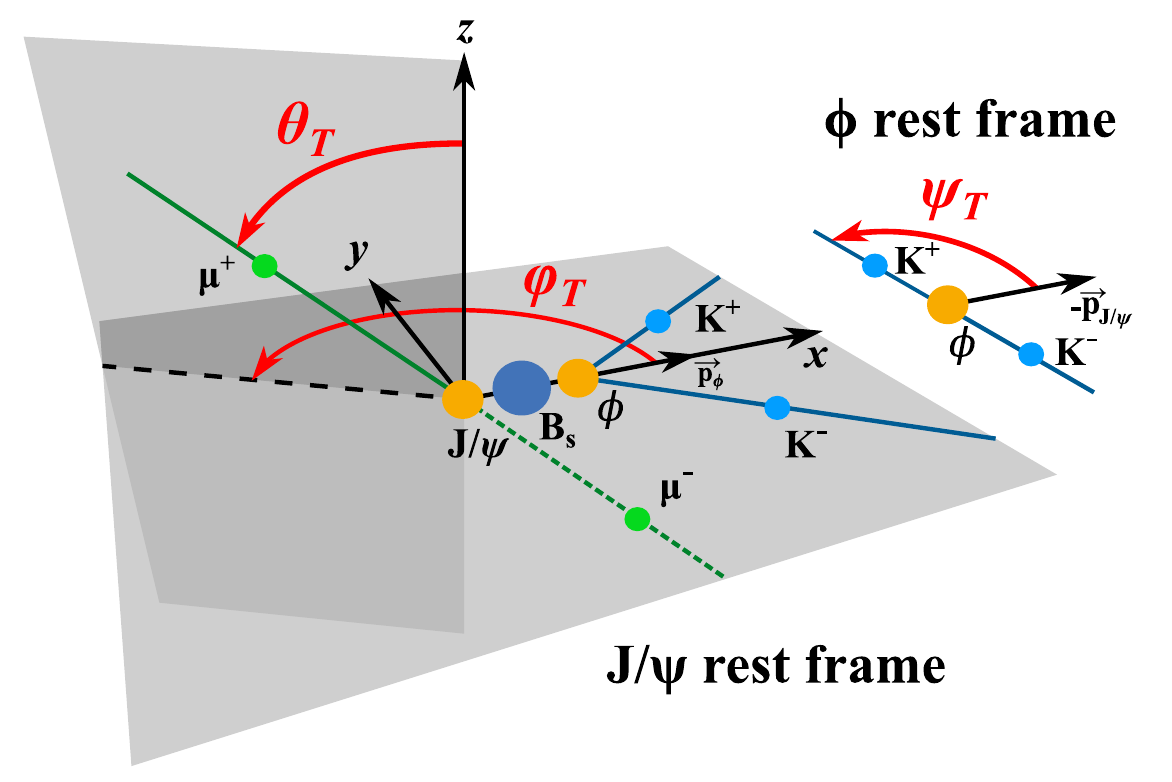}
  \caption{Definition of the three angles \theT, \psiT, and \phiT\ describing the topology of the \BsJpsiphiMMKK decay.}
  \label{fig:angles}
\end{figure}

In this measurement, we use the transversity basis~\cite{ref:Dighe99epjc} defined by the three decay angles $\Theta = (\theT, \psiT, \phiT)$, as illustrated in Fig.~\ref{fig:angles}.
The angles \theT\ and \phiT\ are, respectively, the polar and azimuthal angles of the \PGmp in the rest frame of the \PJGy meson, where the $x$ axis is defined by the direction of the \PGf\ meson momentum and the $x$-$y$ plane is defined by the plane of the $\PGf \to \KK$ decay.
The helicity angle $\psiT$ is the angle of the \PKp meson momentum in the \PGf meson rest frame with respect to the negative \PJGy meson momentum direction.

The differential decay rate of \BsJpsiphiMMKK is described by a function $\mathcal{F}(\Theta,ct,\alpha)$, as in Ref.~\cite{ref:Dighe95plb}:
\begin{linenomath}
\begin{equation}
  \frac{\rd^4\Gamma\!\left(\PBzs\right)}
       {\rd\Theta\,\rd\!\left(ct\right)} =
  \mathcal{F}(\Theta,ct,\alpha)\propto \sum^{10}_{i=1}O_i(ct,\alpha)\, g_i(\Theta),
  \label{eqn:decayrate}
\end{equation}
\end{linenomath}
where $O_i$ are time-dependent functions, $g_i$ are angular functions, and $\alpha$ is a set of physics parameters.

The functions $O_i(ct, \alpha)$ are:
\begin{linenomath}
\ifthenelse{\boolean{cms@external}}
{
\begin{multline}
  O_i(ct, \alpha) = N_i \re^{- \Gs t}\Biggl[ a_i\cosh\left(\frac{\DGs t}{2}\right)+b_i\sinh\left(\frac{\DGs t}{2}\right)    \\
    +c_i\cos(\Dms t)+ d_i\sin(\Dms t)\Biggr],
\label{eqn:observables}
\end{multline}
}
{
\begin{equation}
  O_i(ct, \alpha) =
  N_i \re^{- \Gs t}\left[
    a_i\cosh\left(\frac{\DGs t}{2}\right)+
    b_i\sinh\left(\frac{\DGs t}{2}\right)+
    c_i\cos(\Dms t)+
    d_i\sin(\Dms t)\right],
\label{eqn:observables}
\end{equation}
}
\end{linenomath}
where \Dms (\DGs) is the absolute mass (decay width) difference between the \Bsl and \Bsh mass eigenstates, and \Gs is the average decay width, defined as the arithmetic average of the \Bsl and \Bsh decay widths.
The functions $g_i(\Theta)$ and the parameters $N_i$, $a_i$, $b_i$, $c_i$, and $d_i$ are defined in Table~\ref{tab:kinematics}.

\begin{table*}[h!tb]
  \topcaption{Angular and time-dependent terms of the signal model.}
  \centering
  \renewcommand{\arraystretch}{1.15}
  \cmsTable{
    \begin{tabular}{
        @{\hskip 0pt} l 
        @{\hskip 4pt} c
        @{\hskip 4pt}   
        @{\hskip 4pt} c
        @{\hskip 4pt}   
        @{\hskip 4pt} c
        @{\hskip 4pt}   
        @{\hskip 4pt} c
        @{\hskip 4pt}   
        @{\hskip 4pt} c
        @{\hskip 4pt}   
        @{\hskip 4pt} c
        @{\hskip 0pt}}
      $i$ & $g_i(\theT,\psiT,\phiT)$ &
            $N_i$ & $a_i$ & $b_i$ & $c_i$ & $d_i$ \\
      \hline
      \rule{0pt}{3ex}
      \hspace{-0.2cm} 1 &
      $2\sqcos\psiT(1-\sqsin\theT\sqcos\phiT)$                 &
      $\sqabs{\Azero(0)}$                                      &
      1                                                        &
      $D$                                                      &
      $C$                                                      &
      \hspace{-0.8em}$-S$\\
      2 &
      $\sqsin\psiT(1-\sqsin\theT\sqsin\phiT)$                  &
      $\sqabs{\Apara(0)}$                                      &
      1                                                        &
      $D$                                                      &
      $C$                                                      &
      \hspace{-0.8em}$-S$\\
      3 &
      $\sqsin\psiT\sqsin\theT$                                 &
      $\sqabs{\Aperp(0)}$                                      &
      1                                                        &
      \hspace{-0.8em}$-D$                                      &
      $C$                                                      &
      $S$\\
      4 &
      $-\sqsin\psiT\sin2\theT\sin\phiT$                        &
      $\abs{\Apara(0)}\abs{\Aperp(0)}$                         &
      $C\sin(\dperp-\dpara)$                                   &
      $S\cos(\dperp-\dpara)$                                   &
      $\sin(\dperp-\dpara)$                                    &
      $D\cos(\dperp-\dpara)$\\
      5 &
      $\frac{1}{\sqrt{2}}\sin2\psiT\sqsin\theT\sin2\phiT$      &
      $\abs{\Azero(0)}\abs{\Apara(0)}$                         &
      $\cos(\dpara-\dzero)$                                    &
      $D\cos(\dpara-\dzero)$                                   &
      $C\cos(\dpara-\dzero)$                                   &
      $-S\cos(\dpara-\dzero)$\\
      6 &
      $\frac{1}{\sqrt{2}}\sin2\psiT\sin2\theT\cos\phiT$        &
      $\abs{\Azero(0)}\abs{\Aperp(0)}$                         &
      $C\sin(\dperp-\dzero)$                                   &
      $S\cos(\dperp-\dzero)$                                   &
      $\sin(\dperp-\dzero)$                                    &
      $D\cos(\dperp-\dzero)$\\
      7 &
      $\frac{2}{3}(1-\sqsin\theT\sqcos\phiT)$                  &
      $\sqabs{\Aswav(0)}$                                      &
      1                                                        &
      \hspace{-0.8em}$-D$                                      &
      $C$                                                      &
      $S$\\
      8 &
      $\frac{1}{3}\sqrt{6}\sin\psiT\sqsin\theT\sin2\phiT$      &
      $\abs{\Aswav(0)}\abs{\Apara(0)}$                         &
      $C\cos(\dpara-\dswav)$                                   &
      $S\sin(\dpara-\dswav)$                                   &
      $\cos(\dpara-\dswav)$                                    &
      $D\sin(\dpara-\dswav)$\\
      9 &
      $\frac{1}{3}\sqrt{6}\sin\psiT\sin2\theT\cos\phiT$        &
      $\abs{\Aswav(0)}\abs{\Aperp(0)}$                         &
      $\sin(\dperp-\dswav)$                                    &
      $-D\sin(\dperp-\dswav)$                                  &
      $C\sin(\dperp-\dswav)$                                   &
      $S\sin(\dperp-\dswav)$\\
      10 &
      $\frac{4}{3}\sqrt{3}\cos\psiT(1-\sqsin\theT\sqcos\phiT)$ &
      $\abs{\Aswav(0)}\abs{\Azero(0)}$                         &
      $C\cos(\dzero-\dswav)$                                   &
      $S\sin(\dzero-\dswav)$                                   &
      $\cos(\dzero-\dswav)$                                    &
      $D\sin(\dzero-\dswav)$
    \end{tabular}
  }
  \label{tab:kinematics}
\end{table*}
The coefficients $C$, $S$, and $D$ contain the information about \CP violation, and are defined as:
\begin{linenomath}
\begin{equation*}
  \begin{aligned}
  C & = \frac{1-\sqabs{\lambda}        }{1+\sqabs{\lambda}}, &
  S & =-\frac{2   \abs{\lambda}\sin\phis}{1+\sqabs{\lambda}}, &
  D & =-\frac{2   \abs{\lambda}\cos\phis}{1+\sqabs{\lambda}},
  \end{aligned}
\end{equation*}
\end{linenomath}
using the same sign convention as that in the LHCb measurement~\cite{ref:lhcbAaij13prd}.
The amount of \CP violation in the \PBzs-\PABzs system is given by the complex parameter $\lambda$, defined as $\lambda = (q/p) (\overline{A}_f/A_f)$, where $A_f$ ($\overline{A}_f)$ is the decay amplitude of the \PBzs (\PABzs) meson to the final state $f$, and the parameters $p$ and $q$ relate the mass and flavor eigenstates through $\Bsh = p |\PBzs\rangle - q|\PABzs\rangle$ and $\Bsl = p |\PBzs\rangle + q|\PABzs\rangle$~\cite{ref:Branco99irmp}.
The parameters $\sqabs{\Aperp}$, $\sqabs{\Azero}$, and $\sqabs{\Apara}$ are the magnitudes of the perpendicular, longitudinal, and parallel transversity amplitudes of the \BsJpsiphi decay, respectively; $\sqabs{\Aswav}$ is the magnitude of the $S$-wave amplitude from $\PBzs \to \PJGy\, \PfDzP{980}$ and nonresonant \BsJpsiKK decays, and the parameters $\dperp$, $\dzero$, $\dpara$, and $\dswav$ are the respective strong phases.

Equation~(\ref{eqn:decayrate}) represents the model for the \PBzs meson decay, while the model for the \PABzs meson decay is obtained by changing the sign of the $c_i$ and $d_i$ terms in Eq.~(\ref{eqn:observables}).

\section{The CMS detector}
\label{sec:cmsdet}

The central feature of the CMS apparatus is a superconducting solenoid of 6\unit{m} internal diameter, providing a magnetic field of 3.8\unit{T}. Within the solenoid volume are a silicon pixel and strip tracker, a lead tungstate crystal electromagnetic calorimeter, and a brass and scintillator hadron calorimeter, each composed of a barrel and two endcap sections. Forward calorimeters extend the pseudorapidity ($\eta$) coverage provided by the barrel and endcap detectors. Muons are detected in gas-ionization chambers embedded in the steel flux-return yoke outside the solenoid.

The silicon tracker measures charged particles within the 
range $\abs{\eta} < 2.5$. During the LHC running period when the data used
in this Letter were recorded, the silicon tracker consisted of 1856 silicon
pixel and 15\,148 silicon strip detector modules.

Muons are measured in the range $\abs{\eta} < 2.4$, with detection planes made using three technologies: drift tubes, cathode strip chambers, and resistive plate chambers. The efficiency to reconstruct and identify muons is greater than 96\%. Matching muons to tracks measured in the silicon tracker results in a relative transverse momentum (\pt) resolution, for muons with \pt up to 100\GeV, of 1\% in the barrel and 3\% in the endcaps~\cite{ref:cmsmpr18}.

Events of interest are selected using a two-tiered trigger
system~\cite{ref:cmstri17}. The first level (L1), composed of
custom hardware processors, uses information from the calorimeters and
muon detectors to select events at a rate of around $100\unit{kHz}$ within
a fixed time interval of less than $4\mus$. The second level, known as the
high-level trigger (HLT), consists of a farm of processors running a version
of the full event reconstruction software optimized for fast processing, and
reduces the event rate to around $1\unit{kHz}$ before data storage.

A more detailed description of the CMS detector, together with a definition
of the coordinate system used and the relevant kinematic variables, can be
found in Ref.~\cite{ref:cmsexp08}.

\section{Event selection and simulated samples}
\label{sec:events}

The analysis is performed using data collected in proton-proton ($\Pp\Pp$) collisions at $\sqrt{s} =13\TeV$ during 2017--2018, corresponding to an integrated luminosity of \intL.
A trigger optimized for the detection of \PQb hadrons decaying to \PJGy mesons, along with an additional muon potentially usable for flavor tagging, is used to collect the data sample for the analysis.
At L1, the trigger requires three muons, with the minimum \pt requirement on the highest \pt (leading, $\PGm_1$) and second-highest \pt (subleading, $\PGm_2$) muons of $\pt > 5$ and 3\GeV, respectively, and the dimuon invariant mass $m_{\PGm_1\PGm_2} < 9\GeV$.
There is no \pt requirement on the third muon at L1.
At the HLT, the three muons are required to be within the CMS geometrical acceptance $\abs{\eta} < 2.5$; two of these muons must be oppositely charged, each have $\pt > 3.5\GeV$, form a \PJGy candidate with an invariant mass in the range 2.95--3.25\GeV, and have a probability to originate from a common vertex larger than 0.5\%.
The third muon is required to have $\pt > 2\GeV$ and can be used to infer the flavor of the \PBzs meson at production (\ie, its particle/antiparticle state), exploiting semileptonic $\PQb \to \PGmm + \PX$ decays, as discussed further in Section~\ref{sec:fltagg}.

Additional selection criteria are applied to events passing the HLT requirements.
The numerical values of the selection cuts have been optimized with the help of the \textsc{tmva} package~\cite{ref:Antcheva09cpc, ref:tmvaHocker07pos}, using a genetic algorithm, to maximize the signal purity.
First, \PJGy\ meson candidates are constructed using pairs of opposite-sign muons with $\pt > 3.5\GeV$ and $\abs{\eta} < 2.4$, and compatible with originating from a common vertex, obtained from a Kalman fit~\cite{ref:Fruhwirth87nim}.
Candidates are accepted only if their invariant mass is within $150\MeV$ of the world-average \PJGy meson mass~\cite{ref:pdgTanabashi18prd}.
Next, pairs of opposite-sign tracks satisfying the high-purity requirement~\cite{Chatrchyan:2014fea} with $\pt > 1.2\GeV$ and $\abs{\eta} < 2.5$, not associated with the muons that form the \PJGy candidate, are used to form \PGf candidates.
The \PGf candidates are selected if the track pair has an invariant mass, assuming the kaon mass for both particles, within $10\MeV$ of the world-average \PGf meson mass~\cite{ref:pdgTanabashi18prd}.
Finally, the \PJGy and \PGf candidates are combined to form \PBzs candidates: a common vertex (``\PBzs vertex") is obtained from a fit with the four tracks, two for muons and two for kaons.
The invariant mass of the \PBzs candidate is obtained from a kinematic fit, where the invariant mass of the two muons is constrained to the world-average \PJGy meson mass~\cite{ref:pdgTanabashi18prd}. The mass of the \PGf candidate is not constrained since its natural width exceeds the mass resolution.

Due to the high instantaneous luminosity of proton-proton collisions at the LHC, several primary vertices (PVs) are reconstructed in each event.
The vertex that minimizes the angle between the \PBzs candidate momentum vector and the line connecting this vertex with the \PBzs decay vertex is chosen as the production vertex and is used to determine the characteristics of the \PBzs candidate, such as proper decay length.
We used simulations to study if the PV selection procedure introduces any bias in the measurement.
It was found that in about 97\% of the events, the selected PV is also the closest one to the point of origin of the \PBzs meson.
The impact of choosing a different vertex in the remaining cases on the final results is found to be negligible with respect to the total systematic uncertainties discussed in Section~\ref{sec:result}.
The proper decay length is measured as $\ct = c m^\text{PDG}_\PBzs L_\text{xy} /\pt$, where $m^\text{PDG}_\PBzs$ is the world-average \PBzs mass~\cite{ref:pdgTanabashi18prd} and $L_\text{xy}$ is the reconstructed transverse decay length, which is defined as the distance in the transverse plane from the production vertex to the \PBzs vertex.
Additional selection criteria are applied to \PBzs candidates, requiring $\pt > 11\GeV$, the four-track vertex fit $\chi^2$ probability $ > 2\%$, an invariant mass in the 5.24--5.49\GeV range, and a proper decay length $\ct > 70\mum$, with an uncertainty $\sct < 50\mum$.
The proper decay length uncertainty is obtained by propagating the uncertainties in the decay distance and the \pt of the \PBzs candidate to \ct.
In about 2\% of the events more than one \PBzs candidate is selected.
In these cases, the candidate with the highest vertex fit probability is chosen.
The impact of this choice on the measurement has been evaluated by redoing the analysis using the candidate with the lowest vertex fit probability.
No sizable bias has been observed with respect to the total systematic uncertainties discussed in Section~\ref{sec:result}.
A total of \Nevtot \BsJpsiphi candidates are selected.

Simulated event samples are used to measure the selection efficiency and the flavor tagging performance.
These samples are produced using the \PYTHIA 8.230 Monte Carlo (MC) event generator~\cite{Sjostrand:2014zea} with the underlying event tune CP5~\cite{Sirunyan:2019dfx} and the parton distribution function set NNPDF3.1~\cite{Ball:2017nwa}.
The \PQb hadron decays are modeled with the \EVTGEN 1.6.0 package~\cite{ref:Lange01nim}.
Final-state photon radiation is accounted for in the \EVTGEN simulation with \PHOTOS 215.5~\cite{ref:Barberio91cpc, ref:Barberio94cpc}.
The response of the CMS detector is simulated using the \GEANTfour package~\cite{ref:Agostinelli03nim}. The effect of multiple collisions in the same or neighboring bunch crossings (pileup) is accounted for by overlaying simulated minimum bias events on the hard-scattering process.
Simulated samples are then reconstructed using the same software as for collision data.

\begin{figure}[h!]
  \centering
  \includegraphics[width=0.45\textwidth]{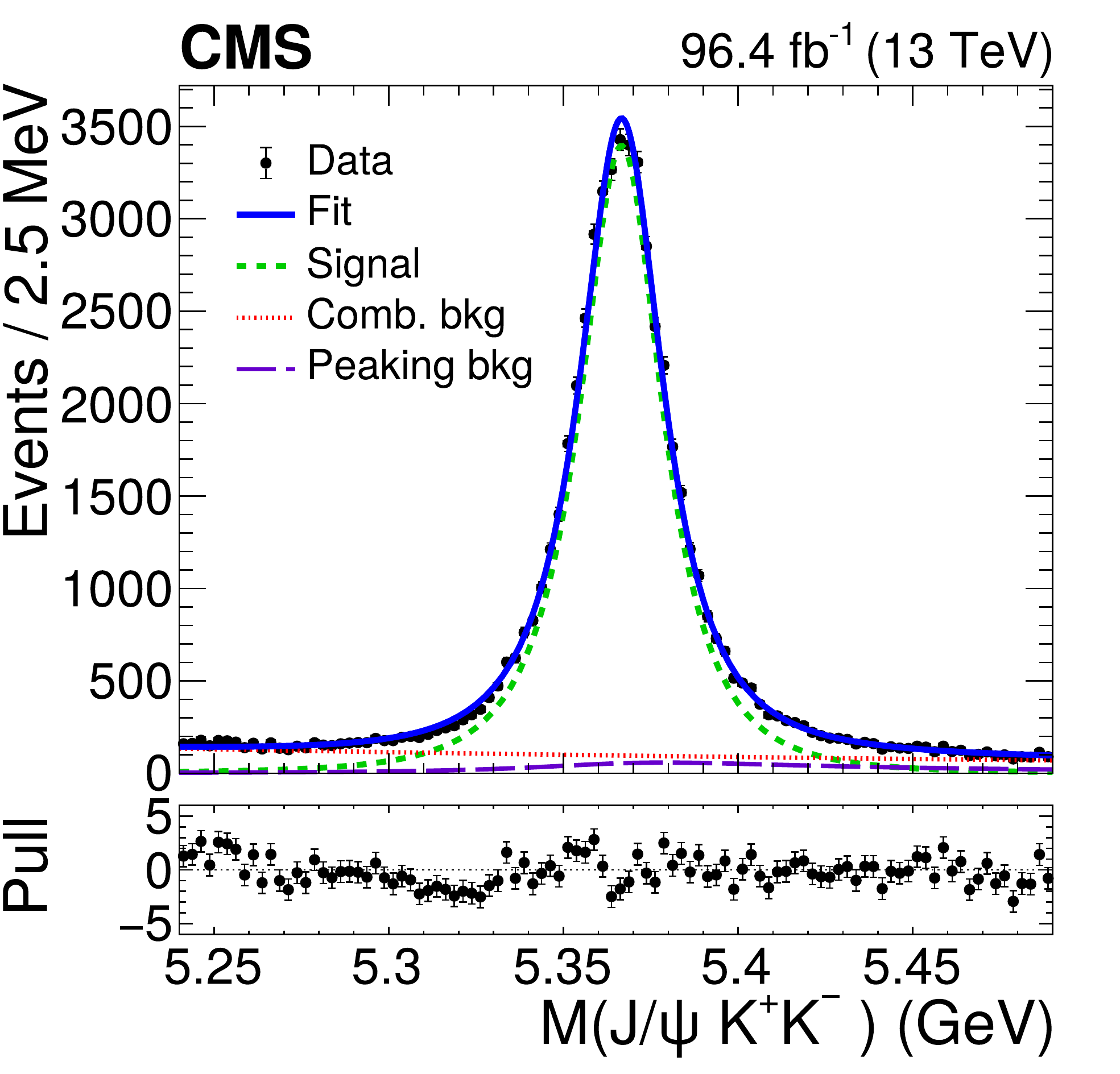}
  \caption{
    The invariant mass distribution of the \BsJpsiphiMMKK candidates in data. 
    The vertical bars on the points represent the statistical uncertainties.
    The solid line represents a projection of the fit to data (as discussed in Section~\protect\ref{sec:mlhfit}, solid markers), the dashed line corresponds to the signal, the dotted line to the combinatorial background, and the long-dashed line to the peaking background from $\PBz \to \PJGy\, \kstar \to \mumu\,\PKp\PGpm$, as obtained from the fit.
    The distribution of the differences between the data and the fit, divided by the combined uncertainty in the data and the best fit function for each bin (pulls) is displayed in the lower panel.}
  \label{fig:massFit}
\end{figure}

\begin{figure}[htb!]
  \centering
  \includegraphics[width=0.45\textwidth]{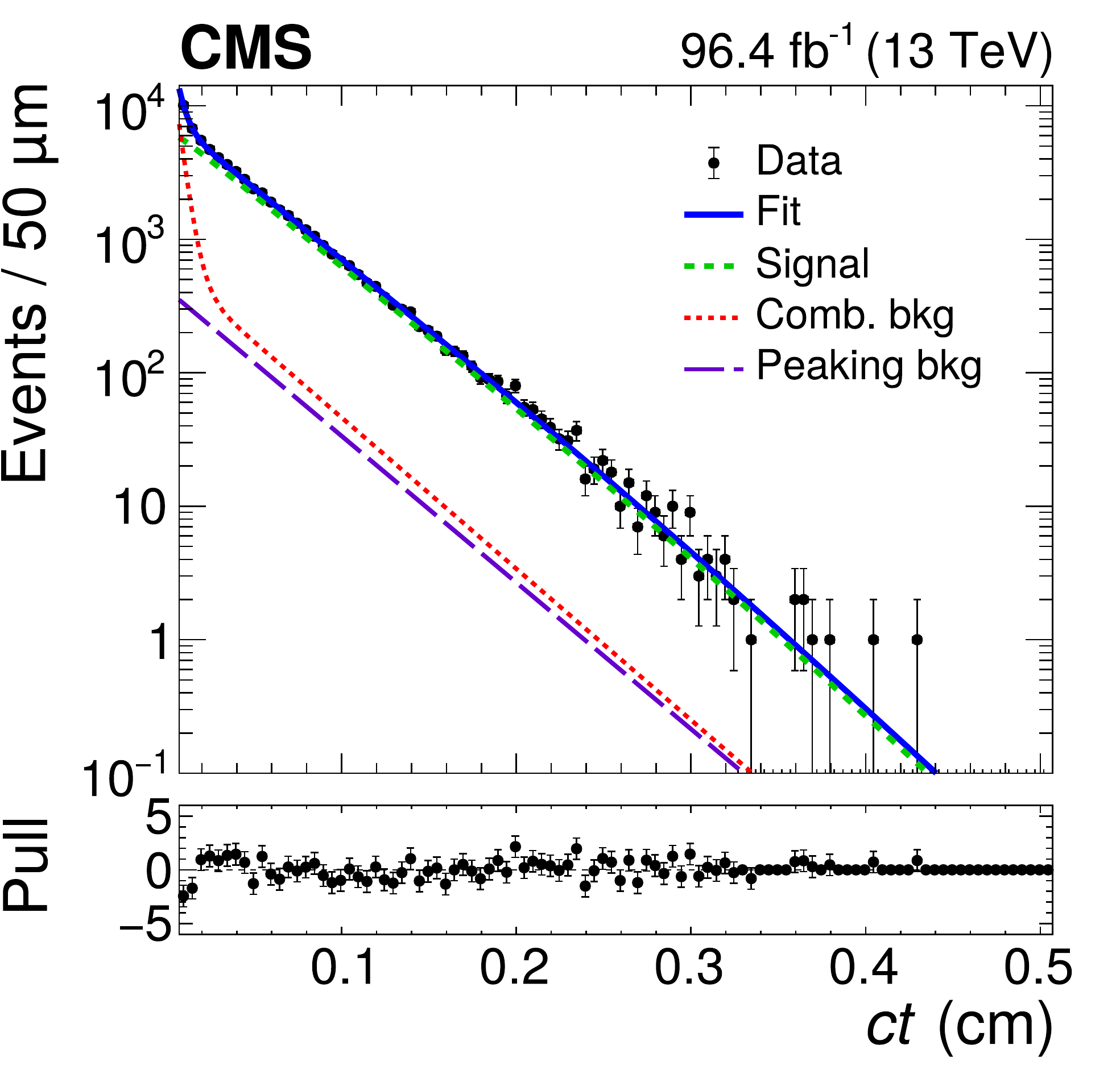}
  \includegraphics[width=0.45\textwidth]{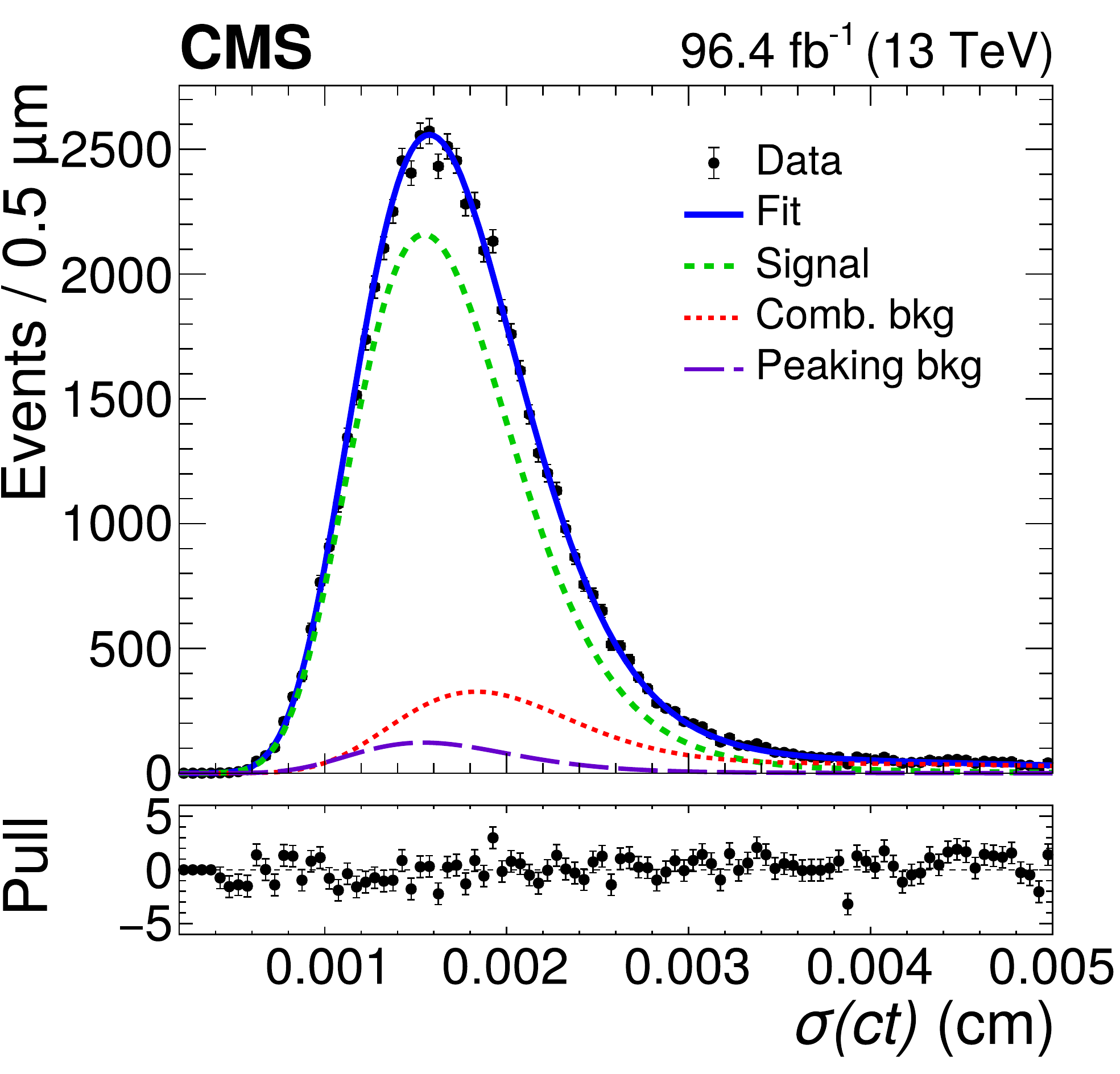}
  \caption{The \ct distribution (\cmsLeft) and its uncertainty (\cmsRight) for the \BsJpsiphiMMKK candidates in data. The notations are as in Fig.~\protect\ref{fig:massFit}.}
  \label{fig:ctimeFit}
\end{figure}

The simulation is validated via comparison with background-subtracted data in a number of control distributions.
The \PBzs candidate invariant mass distribution after the signal selection is shown in Fig.~\ref{fig:massFit}, whereas the proper decay length and its uncertainty distributions are shown in Fig.~\ref{fig:ctimeFit}.


\section{Flavor tagging}
\label{sec:fltagg}
The flavor of the \PBzs candidate at production is determined with an OS flavor tagging algorithm.
The OS approach is based on the fact that \PQb quarks are predominantly produced in $\bbbar$ pairs, and therefore one can infer the initial \PBzs meson flavor by determining the flavor of the other (``OS") \PQb quark in the event.

In this analysis, the flavor of the OS \PQb hadron is deduced by exploiting the semileptonic $\PQb \to \PGmm + \PX$ decay, where the muon sign $\xi$ is used as the tagging variable ($\xi = -1$ for \PBzs).
This technique works on a probabilistic basis.
If no OS muon is found, the event is considered as untagged ($\xi =0$).
The tagging efficiency \etag is defined as the fraction of candidate events that are tagged.
When a muon is found, the tag is defined to be correct (``right tag") if the flavor predicted using the muon sign and the actual \PBzs meson flavor at production coincide.
The correlation between the muon sign and the signal \PBzs meson flavor is diluted by wrong tags (mistags) originating from cascade $\PQb \to\PQc\to\PGmp + \PX$ decays, oscillation of the OS \PBz or \PBzs meson, and muons originating from other sources, such as \PJGy meson and charged pion and kaon decays.
The mistag fraction \wtag is defined as the ratio between the number of wrongly tagged events and the total number of tagged events. It is used to compute the dilution $\D \equiv 1 - 2\wtag$, which is a measure of the performance degradation due to mistagged events.
The tagging power $\Ptag \equiv \etag\D^2$ is the effective tagging efficiency, which takes into account the dilution and is used as a figure of merit in maximizing the algorithm performance.

To maximize the sensitivity of this measurement, we have developed a novel OS muon tagger taking advantage of machine learning techniques.
The use of deep neural networks (DNNs) in the new tagger leads to lowering of the mistag probability \wtag and reducing of the related systematic uncertainties.
The use of a dedicated trigger, which requires an OS muon, dramatically increases the fraction of tagged candidates compared to our earlier measurement~\cite{ref:cmsKhachatryan16plb}.
Taken together, these two improvements increase the muon tagging performance by ${\approx}$20\% compared to that in Ref.~\cite{ref:cmsKhachatryan16plb}.

For each event, we search for a candidate OS muon consistent with originating from the same production vertex as the signal \PBzs meson.
This \textit{tagging} muon is required to have $\pt > 2\GeV$, $\abs{\eta}<2.4$, the longitudinal impact parameter with respect to the production vertex $\mathrm{IP}_z < 1.0\unit{cm}$, and the distance from the \PBzs candidate momenta in the $(\eta,\phi)$ plane $\Delta R_{\eta,\phi}>0.4$.
Tracks that belong to the reconstructed \BsJpsiphiMMKK decay are explicitly excluded from consideration.
In order to reduce the contamination from light-flavor hadrons misreconstructed as tagging muons, a discriminator based on a DNN was developed using the \textsc{Keras} library~\cite{ref:chollet2015keras} within the \textsc{tmva} toolkit.
This discriminator, called the ``DNN against light hadrons'' in the following, uses 25 input features related to the muon kinematics and reconstruction quality, and is trained with $3.5 \times 10^6$ simulated muon candidates of which $2.5 \times 10^5$ are misreconstructed hadrons.
The following DNN hyperparameters are optimized through a grid scan to maximize the discrimination power: number of layers, number of neurons for each layer, and the dropout probability.
No signs of overtraining are observed at the chosen hyperparameters configuration when comparing the output distributions from the testing and training samples.
Tagging muons are required to pass a working point of the DNN output that has an efficiency of ${\approx}$98\% for genuine muons and ${\approx}$33\% for misreconstructed light-flavor hadrons, when evaluated using muon candidates reconstructed with the CMS particle-flow (PF) algorithm~\cite{Sirunyan_2017}.
In ${\approx}$3\% of the events where more than one tagging muon candidate passes all the above selections, only the highest \pt one is kept.

Another DNN is used to further discriminate the right- and wrong-tag muons, as well as to predict the mistag probability on a per-event basis.
This DNN, referred to as the muon tagger DNN, has been developed using the \textsc{Keras} library within the \textsc{tmva} toolkit, based on simulated \BsJpsiphiMMKK events, and calibrated with self-tagging \BJpsiK MC and data samples, as described below.

The input features of the muon tagger DNN are of two kinds: muon variables and cone variables.
The muon variables are the muon \pt, $\eta$, transverse and longitudinal impact parameters with respect to the production vertex, along with their uncertainties, the distance $\Delta R_{\eta,\phi}$ to the signal \PBzs candidate, and the discriminant of the DNN against light hadrons.
The cone variables are related to the activity in a cone of radius $\Delta R_{\eta,\,\phi}=0.4$ around the muon momentum direction and include the relative PF isolation~\cite{Sirunyan_2017}, the scalar \pt sum of all additional tracks within the cone, the sum of their charges weighted by the track \pt, the muon relative momentum and $\Delta R_{\eta,\,\phi}$ with respect to the vector sum of the momenta of all additional tracks within the cone, and the ratio of the energy of the muon to the total energy of all additional tracks within the cone (assuming the pion mass for each track).
The muon tagger DNN is trained on $2.8 \times 10^5$ simulated \BsJpsiphi events, of which about $85\,000$ have a wrong tag.
Its structure is optimized similarly to that for the DNN against light hadrons.
The optimal DNN has three dense layers of 200 neurons, each with a rectified linear unit activation function.
A dropout layer with a dropout probability of 40\% is placed after each dense layer.
The cross-entropy loss function and the Adam optimizer~\cite{kingma2014adam} are used.
The DNN is constructed in such a way that its output score $d$ is equal to the probability of tagging the event correctly.
Therefore, the per-event mistag probability is simply $\wevt = 1 - d$.

The output $d$ of the tagger is calibrated using a self-tagging data sample of \BJpsiKMM events, where the charge of the kaon corresponds to the charge and flavor of the \PBpm meson.
The same trigger and \PJGy candidate reconstruction requirements as for the \PBzs signal sample are applied.
A charged particle with $\pt> 1.6\GeV$, assumed to be a kaon, is combined in a kinematic fit with the dimuon pair to form the \PBpm candidate.
The calibration is performed separately for the 2017 and 2018 data samples, by comparing the measured mistag fraction (\wmeas) with the \wevt predicted by the muon tagger DNN.
The \PBpm events are divided into 100 bins in \wevt and the right- and wrong-tag events are separately counted in each bin to extract the corresponding $\wmeas$ value.
The \PBpm signal in each bin is discriminated from the background via a binned likelihood fit to the $\PJGy\PKpm$ invariant mass distribution in the 5.10--5.65\GeV range.

\begin{figure}[htb!]
  \centering
    \includegraphics[width=0.45\textwidth]{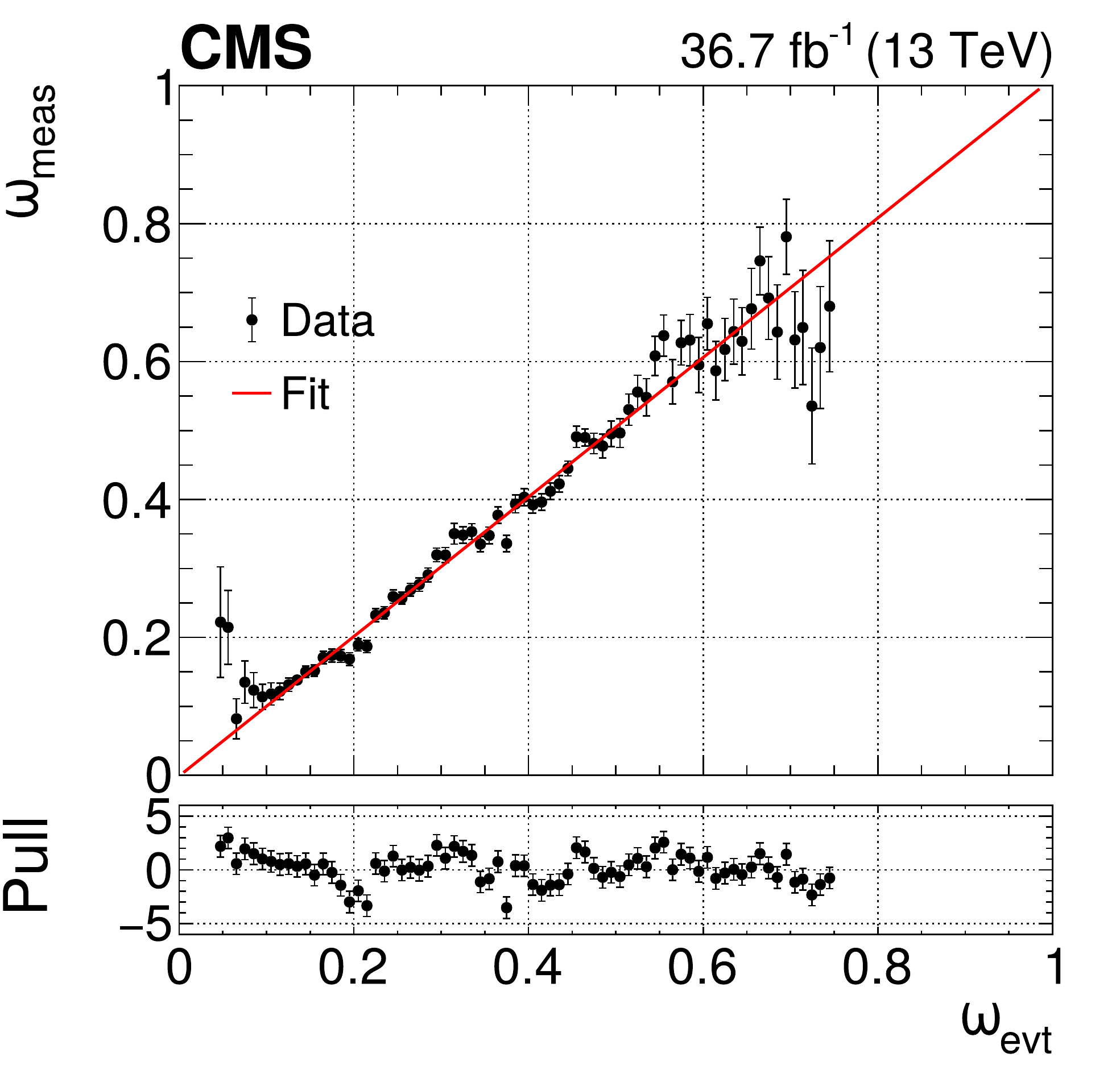}
    \includegraphics[width=0.45\textwidth]{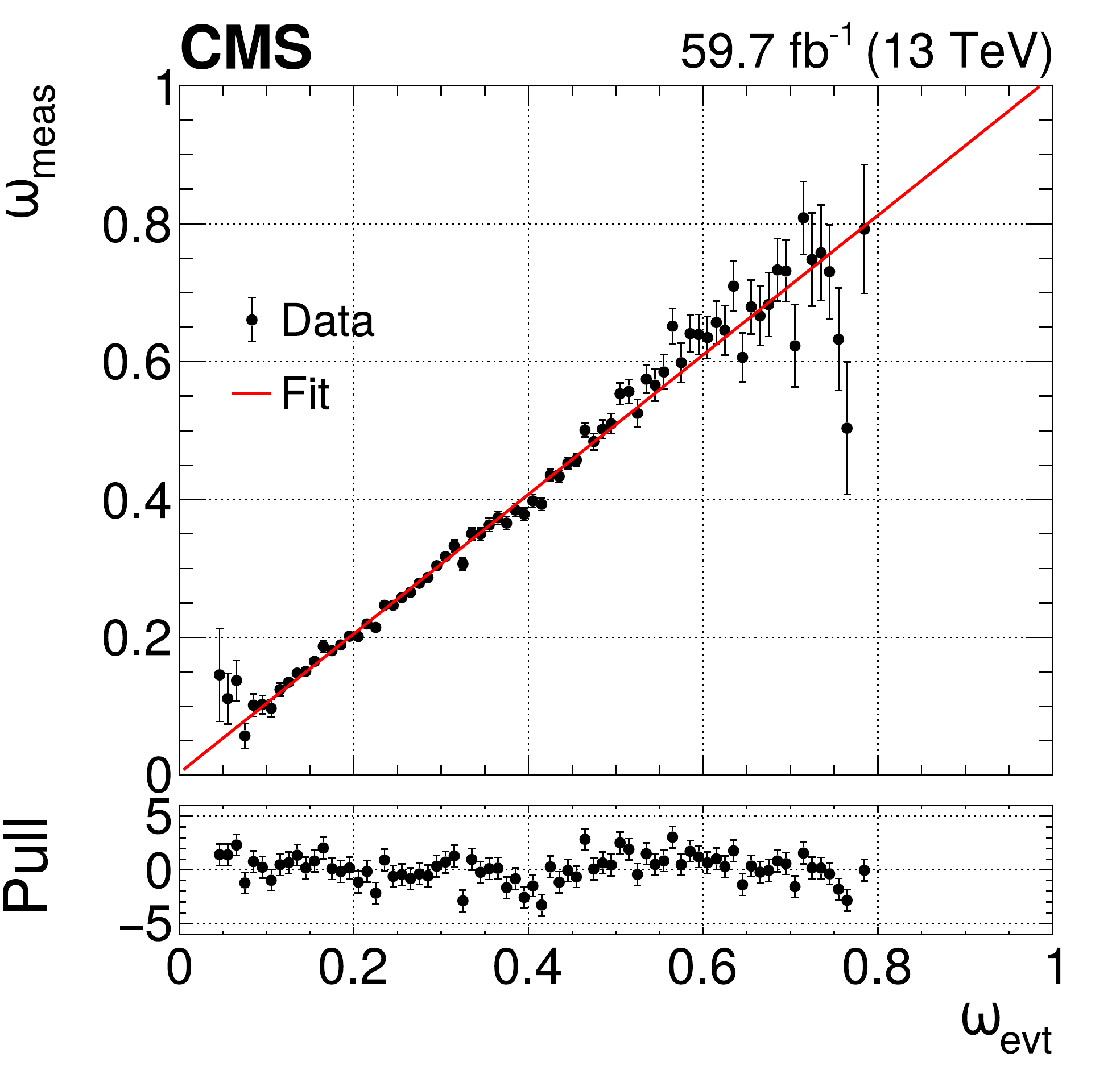}
    \caption{
      Results of the calibration of the per-event mistag probability \wevt based on \BJpsiKMM decays from the 2017 (\cmsLeft) and 2018 (\cmsRight) data samples.
      The vertical bars represent the statistical uncertainties.
      The solid line shows a linear fit to data (solid markers).
      The pull distributions between the data and the fit function in each bin are shown in the lower panels.}
    \label{fig:fltaggcalibration}
\end{figure}

The calibration results for the 2017 and 2018 \PBpm data are shown in Fig.~\ref{fig:fltaggcalibration}.
The data points are fitted with a linear function $a + b\wevt$.
The calibration parameters returned by the fit for the 2017 (2018) data samples are $a=-0.0010\pm0.0040$, $b=1.012\pm0.013$ ($a=0.0031\pm0.0031$, $b=1.011\pm0.010$), statistically compatible with a unit slope and zero offset.

The calibration of the DNN output is also verified with a procedure similar to that described above using an independent sample of simulated \BsJpsiphi and \BJpsiK events.
The reconstructed \PBzs and \PBpm mesons are matched to the generated ones in order to find their true flavor at production.
In general, the measured mistag probability is predicted very accurately by \wevt over the entire measured range for all the examined samples and processes, with more than 90\% of the tagged events falling in the $\wevt =$ 0.1--0.5 range in all cases.
Residual differences are well approximated by linear functions with slopes close to unity and offsets consistent with zero. The $\chi^2$ per degree of freedom values for all fits are below 2.
We conclude that the value of \wevt returned by the tagging DNN is a good approximation of the true mistag probability in data, with minor residual differences taken into account with calibration functions.

The calibrated flavor tagger performance, evaluated using \BJpsiK events in data, is shown in Table~\ref{tab:mutagperformances}.
A tagging efficiency of ${\approx}$50\% and a tagging power of ${\approx}$10\% are achieved in both the 2017 and 2018 data samples.
The efficiency is much higher than the semileptonic \PQb hadron branching fraction due to the requirement of an additional OS muon at the HLT, as described in Section~\ref{sec:events}.

Possible differences in the mistag probability calibration between the \PBzs and \PBpm samples, as well as the statistical uncertainties in the calibration parameters and possible variations from linearity of the calibration function, are considered as systematic uncertainties and described in Section~\ref{sec:result}.

\begin{table}[h]
  \centering
  \topcaption{Calibrated opposite-side muon tagger performance evaluated using \BJpsiK events in the 2017 and 2018 data samples. The uncertainties shown are statistical only.}
  \begin{tabular}{cccc}
    Data sample & $\etag$ (\%) & $\wtag$ (\%) & $\Ptag$ (\%) \\
    \hline
    2017 & $45.7 \pm 0.1$ & $ 27.1 \pm 0.1$ & $\hphantom{1}9.6 \pm 0.1 $ \\
    2018 & $50.9 \pm 0.1$ & $ 27.3\pm 0.1$ & $ 10.5 \pm 0.1 $ \\
  \end{tabular}
  \label{tab:mutagperformances}
\end{table}

\section{Maximum-likelihood fit}
\label{sec:mlhfit}

{\tolerance=800 An unbinned multidimensional extended maximum-likelihood fit is performed on the combined data samples using 8 observables as input: the \PBzs candidate invariant mass $m_\PBzs$, the three decay angles $\Theta$ of the reconstructed \PBzs candidate, the flavor tag decision $\xi$, the mistag fraction \wevt, the proper decay length of the \PBzs candidate $\ct$, and its uncertainty $\sct$. \par}

From the multidimensional fit, the physics parameters of interest \phis, \DGs, \Gs, \Dms, $\abs{\lambda}$, the squares of amplitudes $\abs{\Aperp}^2$, $\abs{\Azero}^2$, $\abs{\Aswav}^2$, and the strong phases $\dpara$, $\dperp$, and $\dswpd$ are determined, where $\dswpd$ is defined as the difference $\dswav-\dperp$.
The \BsJpsiphi amplitudes are normalized to unity by constraining $\abs{\Apara}^2$ to $1-\abs{\Aperp}^2-\abs{\Azero}^2$.
The fit model is validated with simulated pseudo-experiments and with simulated samples with different input parameter sets.

The likelihood function is composed of the probability density functions (pdfs) describing the signal and background components.
The likelihood fit algorithm is implemented using the \textsc{RooFit} package~\cite{ref:Antcheva09cpc,Verkerke:2003ir}.
The signal and background pdfs are formed as the product of functions that model the invariant mass distribution and the time-dependent decay rates of the reconstructed candidates.
In addition, the signal pdf includes the efficiency functions.
The event pdf $P$ is defined as:
\begin{linenomath}
\begin{equation}
  P = \frac{N_\text{sig}}{N_\text{tot}}\,P_\text{sig} + \frac{N_\text{bkg}}{N_\text{tot}}\,P_\text{bkg},
  \label{eq:fitmodel}
\end{equation}
\end{linenomath}
where
\begin{linenomath}
\ifthenelse{\boolean{cms@external}}
{
\begin{multline}
  P_\text{sig} = \varepsilon(\ct)\,\varepsilon(\Theta)\,[\widetilde{\mathcal{F}}(\Theta,\ct,\alpha) \otimes G(\ct,\sct) ] \\
  \times P_\text{sig}(m_\PBzs)\, P_\text{sig}(\sct)\,P_\text{sig}(\xi)
\label{eq:fitsgn}
\end{multline}
} 
{
\begin{equation}
  P_\text{sig} = \varepsilon(\ct)\,\varepsilon(\Theta)\,[\widetilde{\mathcal{F}}(\Theta,\ct,\alpha) \otimes G(\ct,\sct) ]\,P_\text{sig}(m_\PBzs)\, P_\text{sig}(\sct)\,P_\text{sig}(\xi)
\label{eq:fitsgn}
\end{equation}
}
\end{linenomath}
and
\begin{linenomath}
\ifthenelse{\boolean{cms@external}}
{
\begin{multline}
  P_\text{bkg} = P_\text{bkg}(\cos{\theta_T},\phi_T)\,P_\text{bkg}(\cos{\psiT})\\
  \times P_\text{bkg}(\ct)\, P_\text{bkg}(m_\PBzs)\, P_\text{bkg}(\sct)P_\text{bkg}(\xi).
  \label{eq:fitbkg}
\end{multline}
}
{
\begin{equation}
  P_\text{bkg} = P_\text{bkg}(\cos{\theta_T},\phi_T)\,P_\text{bkg}(\cos{\psiT})\, P_\text{bkg}(\ct)\, P_\text{bkg}(m_\PBzs)\, P_\text{bkg}(\sct)P_\text{bkg}(\xi).
  \label{eq:fitbkg}
\end{equation}
}
\end{linenomath}
The corresponding negative log likelihood is:
\begin{linenomath}
\begin{equation}
  -\ln{\mathcal{L}} = -\sum^{N_\text{evt}}_{i=0} \ln{P_i} + N_\text{tot} - N_\text{evt}\ln{N_\text{tot}}.
\end{equation}
\end{linenomath}
Here, $P_\text{sig}$ and $P_\text{bkg}$ are the pdfs that describe the \BsJpsiphiMMKK signal and background contributions, respectively.
The yields of signal and background events are $N_\text{sig}$ and $N_\text{bkg}$, respectively, $N_\text{tot}$ is their sum, and $N_\text{evt} = \Nevtot$ is the number of candidates selected in data.
The pdf $\widetilde{\mathcal{F}}(\Theta,\ct,\alpha)$ is the differential decay rate function $\mathcal{F}(\Theta,\ct,\alpha)$ defined in Eq.~(\ref{eqn:decayrate}), modified to include the flavor information $\xi$ and the dilution term $(1-2\wevt)$, which are applied as multiplicative factors to each of the $c_i$ and $d_i$ terms in Eq.~(\ref{eqn:observables}).
In the $\widetilde{\mathcal{F}}$ expression, the value of $\dzero$ is set to zero, following a general convention~\cite{ref:d0Abazov12prd,ref:cdfAaltonen12prd}, and the value of $\DGs$ is constrained to be positive, based on the LHCb measurement~\cite{ref:Aaij2012eq}.
All the parameters of the pdfs are allowed to float in the final fit, unless explicitly stated otherwise.

The functions $\varepsilon(\ct)$ and $\varepsilon(\Theta)$ model the dependence of the signal reconstruction efficiency on the proper decay length and the three angles of the transversity basis, respectively.
The proper decay length efficiency is parameterized with a fourth-order Chebyshev polynomial multiplied by an exponential function with a negative slope, while the angular efficiency is parameterized with spherical harmonics and Legendre polynomials up to order six.
Both parameterizations are obtained from fits to the respective efficiency histograms in \BsJpsiphi simulated events, and are fixed in the fit to data.

The term $G(\ct,\sct)$ is a Gaussian resolution function, which makes use of the per-event decay length uncertainty $\sct$, scaled by a correction factor $\kappa$ introduced to account for the residual effects when the decay length uncertainty is used to model the $\ct$ resolution.
The value of $\kappa$ is estimated using simulated samples and is equal to ${\approx}$1.2 for both the 2017 and 2018 data samples.

The signal mass pdf $P_\text{sig}(m_\PBzs)$ is a Johnson's $S_{\mathrm{U}}$ distribution~\cite{Johnson:1949zj}, while the decay length uncertainty pdf $P_\text{sig}(\sct)$ is described by the sum of two Gamma distributions.
These pdfs best model each individual variable in one-dimensional fits to simulated samples.

The background pdf contains two terms to model both the combinatorial background and the peaking background, dominated by $\PBz \to \PJGy\, \kstar \to \mumu\,\PKp\PGpm$, where the pion is assumed to be a kaon candidate. The background from $\PGLzb \to \PJGy\,\Pp\PKm\to \mumu\,\Pp\PKm$, where the proton is assumed to be a kaon candidate, is estimated using simulated events to have a negligible effect on the fit results compared to the systematic uncertainties discussed in Section~\ref{sec:result}.
The background invariant mass pdf $P_\text{bkg}(m_\PBzs$) is described by an exponential function for the combinatorial background and a Johnson's $S_{\mathrm{U}}$ distribution for the peaking background.
The background decay length pdf $P_\text{bkg}(\ct)$ is described by the sum of two exponential distributions for the combinatorial background, while a single exponential distribution is used for the peaking background.
The angular parts of the background pdfs $P_\text{bkg}(\cos\theT,\phiT)$ and $P_\text{bkg}(\cos\psiT)$ are described analytically by a series of Legendre polynomials for $\cos\theT$ and $\cos\psiT$, and sinusoidal functions for $\phiT$.
For the $\cos\theT$ and $\phiT$ variables, a two-dimensional pdf is used to take into account a possible correlation between the two.
The background decay length uncertainty pdf $P_\text{bkg}(\sct)$ is described by a sum of two Gamma distributions for the combinatorial background, while the peaking background is fixed to that for the signal.

The tag pdfs are defined as $P(\xi) = 1 - \etag$ for the untagged events ($\xi=0$) and $P(\xi) = \etag (1 \pm A_\text{tag})/2$ for the tagged ones ($\xi=\pm 1$), where $\etag$ is the tagging efficiency and $A_\text{tag}$ is the tagging asymmetry, defined as the difference between the numbers of positively and negatively tagged events ($\xi = \pm 1$) divided by the total number.
The measured tagging asymmetry is found to be compatible with zero.

The correlation between the different fit components has been studied in both data and simulations, and found to be negligible.

The peaking background part of $P_\text{bkg}$ is determined using simulated samples, while the initial combinatorial background part is found from a fit to the \PBzs invariant mass sidebands 5.24--5.28\GeV and 5.45--5.49\GeV in data, and then left free to float in the final fit, starting from this initial pdf.
The signal and background components of the decay length uncertainty pdf are fixed to the ones obtained from a two-dimensional fit together with the invariant mass pdf.
The 2017 and 2018 data samples are fitted simultaneously.
The joint likelihood function of the simultaneous fit shares the decay rate model, the invariant mass pdfs, the peaking background model, and the lifetime and angular components of the combinatorial background model between the two samples.
The number of signal and background events are measured separately in each data sample, as is the tagging efficiency.
The efficiency functions, $P_\mathrm{}(\sct)$ pdfs, tag pdfs, and $\kappa$ factors are also specific to each data sample.

\section{Systematic uncertainties and results}
\label{sec:result}

\begin{figure*}[htb]
  \includegraphics[width=0.31\textwidth]{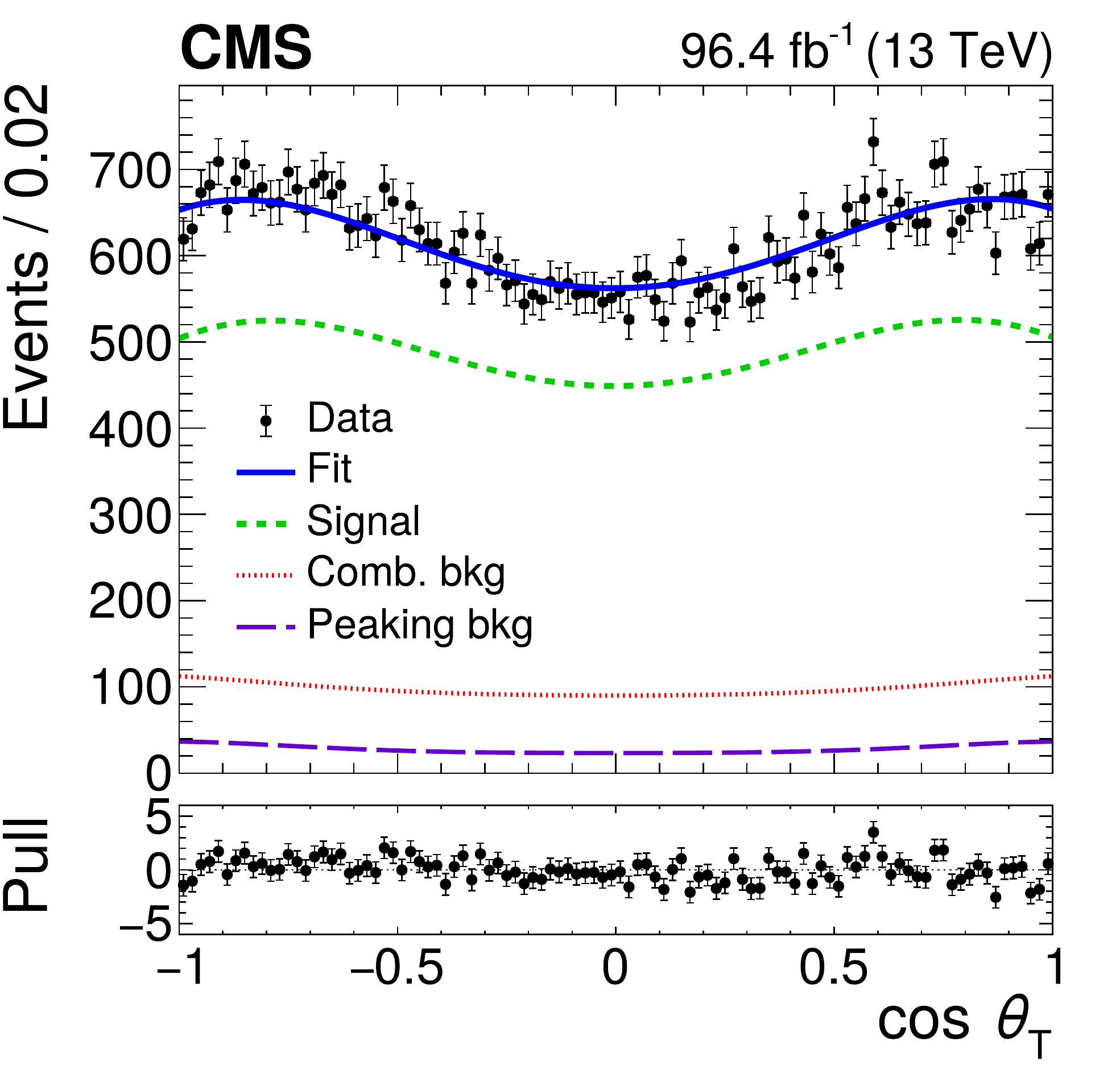} \hfil
  \includegraphics[width=0.31\textwidth]{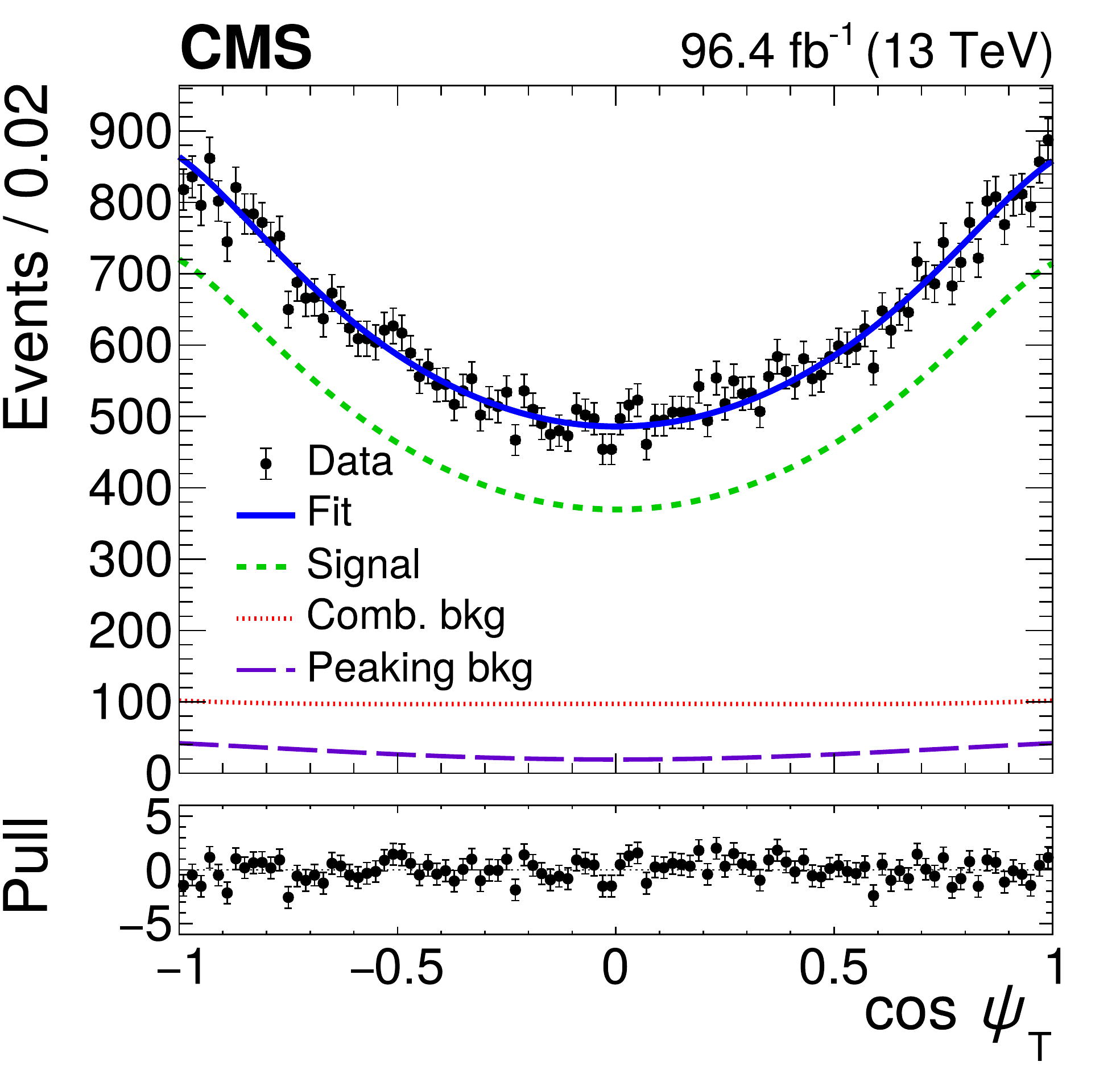} \hfil
  \includegraphics[width=0.31\textwidth]{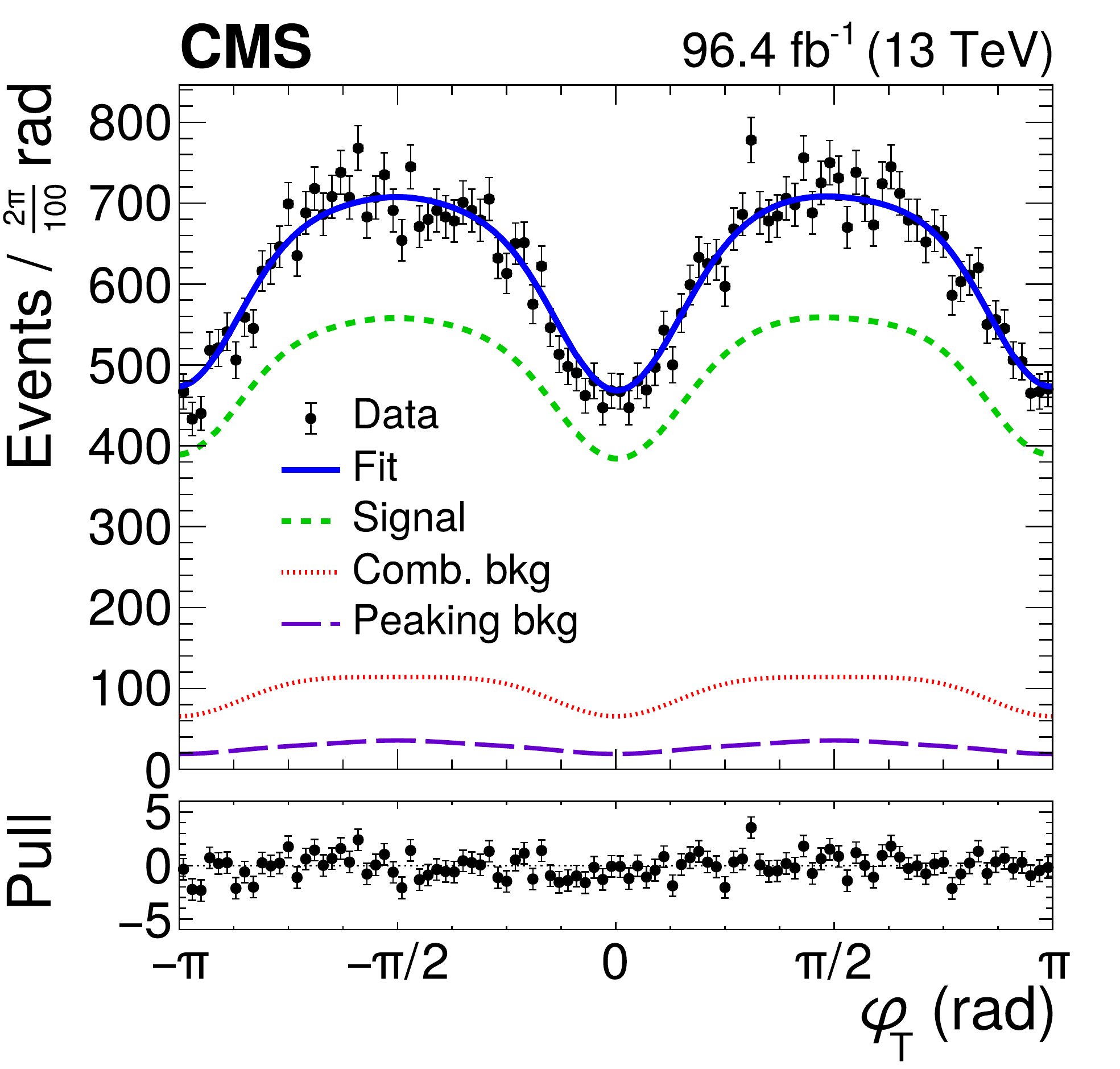}
  \caption{The angular distributions $\cos\theT$ (left), $\cos\psiT$ (middle), and $\phiT$ (right) for the \PBzs candidates and the projections from the fit. The notations are as in Fig.~\protect\ref{fig:massFit}.}
  \label{fig:angleFit}
\end{figure*}

The results of the fit with their statistical and systematic uncertainties are given in Table~\ref{tab:fitresult}, whereas the statistical correlations between the measured parameters are reported in \suppMaterial.
Statistical uncertainties are obtained from the increase in $-\log{\mathcal{L}}$ by 0.5, whereas systematic uncertainties are described below and summarized in Table~\protect\ref{tab:systematics}.
The measured number of \BsJpsiphiMMKK signal events from the fit is $\Nev \pm 250$.
The distributions of the input observables and the corresponding fit projections are shown in Figs.~\ref{fig:massFit}, \ref{fig:ctimeFit}, and \ref{fig:angleFit}.

\begin {table}[htb]
  \centering
  \topcaption{Results of the fit to data. Statistical uncertainties are obtained from the increase in $-\log{\mathcal{L}}$ by 0.5, whereas systematic uncertainties are described below and summarized in Table~\protect\ref{tab:systematics}.}
  \begin{tabular}{ l r@{}l l l}
    Parameter           & \multicolumn{2}{c}{Fit value} & \multicolumn{1}{c}{Stat. uncer.} & \multicolumn{1}{c}{Syst. uncer.} \\
    \hline
    $\phis$~[mrad]      & $-11 $  &         & $\pm\, 50$                & $\pm\, 10 $     \\
    $\DGs$~[ps$^{-1}$]  & $ 0$    & $.114$  & $\pm\, 0.014$             & $\pm\, 0.007 $  \\
    $\Dms~$[\hps]       & $ 17$   & $.51$   & $^{+\,0.10}_{-\,0.09}$    & $\pm\, 0.03 $   \\
    $\abs{\lambda}$     & $ 0$    & $.972$  & $\pm\, 0.026$             & $\pm\, 0.008 $  \\
    $\Gs$~[ps$^{-1}$]   & $ 0 $   & $.6531$ & $\pm\, 0.0042 $           & $\pm\, 0.0026 $ \\
    $\sqabs{\Azero}$    & $ 0 $   & $.5350$ & $\pm\, 0.0047$            & $\pm\, 0.0049 $ \\
    $\sqabs{\Aperp}$    & $ 0 $   & $.2337$ & $\pm\, 0.0063$            & $\pm\, 0.0045 $ \\
    $\sqabs{\Aswav}$    & $ 0 $   & $.022$  & $^{+\,0.008}_{-\,0.007}$  & $\pm\, 0.016 $  \\
    $\dpara$~[rad]      & $ 3 $   & $.18$   & $\pm\, 0.12$              & $\pm\, 0.03 $   \\
    $\dperp$~[rad]      & $ 2 $   & $.77$   & $\pm\, 0.16$              & $\pm\, 0.05 $   \\
    $\dswpd$~[rad]      & $ 0 $   & $.221$  & $^{+\,0.083}_{-\,0.070}$  & $\pm\, 0.048 $  \\
  \end{tabular}
  \label{tab:fitresult}
\end{table}

Several sources of systematic uncertainties in the physics parameters are studied by testing the various assumptions made in the fit model and those associated with the fitting procedure.

\begin{table*}
  \centering
  \topcaption{Summary of the systematic uncertainties.
  The dashes (\NA) mean that the corresponding uncertainty is not applicable.
  The total systematic uncertainty is obtained as the quadratic sum of the individual contributions.}
  \cmsTable{
  \begin{tabular}{lccccccccccc}
    & \phis            &  \DGs            &  \Dms
    & $\abs{\lambda}$  & $\Gs$
    & $\sqabs{\Azero}$ & $\sqabs{\Aperp}$ & $\sqabs{\Aswav}$
    & $\dpara$         & $\dperp$         & $\dswpd$
    \\
    & [mrad]           & [ps$^{-1}$]      & [\hps]
    &                  & [ps$^{-1}$]
    &                  &                  &
    & [rad]            & [rad]            & [rad]
    \\ \hline
    Statistical uncertainty
    & $50$             & $0.014$          & $0.10$
    & $0.026$          & $0.0042$
    & $0.0047$         & $0.0063$         & $0.0077$
    & $0.12$           & $0.16$           & $0.083$
    \\    [\cmsTabSkip]
    Model bias
    & $7.9$            & $0.0019$         & \NA
    & $0.0035$         & $0.0005$
    & $0.0002$         & $0.0012$         & $0.001$
    & $0.020$          & $0.016$          & $0.006$
    \\
    Model assumptions
    & \NA             & \NA               & \NA
    & $0.0046$        & $0.0003$
    & \NA             & $0.0013$          & $0.001$
    & $0.017$         & $0.019$           & $0.011$
    \\
    Angular efficiency
    & $3.8$            & $0.0006$         & $0.007$
    & $0.0057$         & $0.0002$
    & $0.0008$         & $0.0010$         & $0.002$
    & $0.006$          & $0.015$          & $0.015$
    \\
    Proper decay length efficiency
    & $0.3$            & $0.0062$         & $0.001$
    & $0.0002$         & $0.0022$
    & $0.0014$         & $0.0023$         & $0.001$
    & $0.001$          & $0.002$          & $0.002$
    \\
    Proper decay length resolution
    & $3.5$            & $0.0009$         & $0.021$
    & $0.0015$         & $0.0006$
    & $0.0007$         & $0.0009$         & $0.007$
    & $0.006$          & $0.025$          & $0.022$
    \\
    Data/simulation difference
    & $0.6$            & $0.0008$         & $0.004$
    & $0.0003$         & $0.0003$
    & $0.0044$         & $0.0029$         & $0.007$
    & $0.007$          & $0.007$          & $0.028$
    \\
    Flavor tagging
    & $0.5$            & $<$$10^{-4}$     & $0.006$
    & $0.0002$         & $<$$10^{-4}$
    & $0.0003$         & $<$$10^{-4}$     & $<$$10^{-3}$
    & $0.001$          & $0.007$          & $0.001$
    \\
    Sig./bkg. \wevt difference
    & $3.0$           & \NA               & \NA
    & \NA             & $0.0005$
    & \NA             & $0.0008$          & \NA
    & \NA             & \NA               & $0.006$
    \\
    Peaking background
    & $0.3$           & $0.0008$          & $0.011$
    & $<$$10^{-4}$    & $0.0002$
    & $0.0005$        & $0.0002$          & $0.003$
    & $0.005$         & $0.007$           & $0.011$
    \\
    $S$-$P$ wave interference
    & \NA             & $0.0010$          & $0.019$
    & \NA             & $0.0005$
    & $0.0005$        & \NA               & $0.013$
    & \NA             & $0.019$           & $0.019$
    \\
    $P(\sct)$ uncertainty
    & $<$$10^{-1}$    & $0.0019$          & $0.028$
    & $0.0004$        & $0.0008$
    & $0.0006$        & $0.0008$          & $0.001$
    & $0.001$         & $0.002$           & $0.005$
    \\   [\cmsTabSkip]
    Total systematic uncertainty
    & $10.0$           & $0.0070$         & $0.032$
    & $0.0083$         & $0.0026$
    & $0.0049$         & $0.0045$         & $0.016$
    & $0.028$          & $0.045$          & $0.048$
    \\
  \end{tabular}
  }
  \label{tab:systematics}
\end{table*}

  \textit{Model bias:}
  Possible biases in the fitting procedure are evaluated by generating 1000 pseudo-experiments, each statistically equivalent to the data samples, from the fitted model in data (referred to as ``nominal-model pseudo-experiments'' in what follows). 
  Each of them is fitted with the nominal model, and the pull distributions (\ie, the difference divided by the combined uncertainty) between the parameters obtained from the fit and their input values are produced.
  Each pull distribution is fitted with a Gaussian function, and the estimated central value is taken as the corresponding systematic uncertainty, if different from zero by more than its error.
  To avoid double-counting this uncertainty, whenever pseudo-experiments are used to evaluate other systematic uncertainties, the model bias is always subtracted.
  In these cases, the corresponding pull distributions are compared to those obtained with the nominal-model pseudo-experiments. If the mean of the pull distribution differs from the mean of the nominal-model distribution by more than their combined RMS, the difference is taken as the corresponding systematic uncertainty.

  \textit{Model assumptions:}
  The assumptions made in defining the likelihood functions are tested by generating pseudo-experiments with different hypotheses and fitting the samples with the nominal model.
  The following assumptions are tested: signal and background invariant mass models, background proper decay length model, and background angular model.
  Pull distributions with respect to the input values are used to evaluate the systematic uncertainty, as described in the ``model bias'' paragraph.

  \textit{Angular efficiency:}
  The systematic uncertainty related to the limited MC event count used to estimate the angular efficiency function is evaluated by regenerating the efficiency histograms 1000 times using the reference one, with the fit repeated after reestimating the efficiency.
  The root mean square (RMS) of the obtained physics parameter distributions is taken as the systematic uncertainty.

  \textit{Proper decay length efficiency:}
  The proper decay length efficiency is first validated by fitting the \PBpm proper decay length distribution in the control \BJpsiK channel, using several different data-taking periods.
  Each fit is performed applying the efficiency function evaluated using simulated \BJpsiK samples with the same procedure used for the \BsJpsiphi analysis.
  We consider eight different data-taking periods, each with the number of \PBpm candidates comparable with the number of signal candidates in the \PBzs sample used in the analysis.
  We also consider the 2017 and 2018 data-taking periods as two additional large control data sets.
  The results are in good agreement with the world-average \PBpm meson lifetime~\cite{ref:pdgTanabashi18prd}, with differences no larger than 1.5 standard deviations, showing no bias or instabilities during the data taking.
  Having verified that the efficiency parameterization does not introduce any noticeable bias, we evaluate the related systematic uncertainty by varying the parameters of the proper decay length efficiency function within their statistical uncertainties.
  The RMS of the distribution of each extracted physics parameter of interest with respect to the nominal fit value is taken as the corresponding systematic uncertainty.
  We assign a systematic uncertainty to the efficiency model by repeating the fit using the efficiency histogram instead of a smooth efficiency function, and taking the difference from the nominal result as the uncertainty.

  \textit{Proper decay length resolution:}
  A systematic uncertainty is assigned to the proper decay length resolution by varying the $\kappa$ correction factor by ${\pm}10\%$, as estimated from a data-to-simulation comparison, repeating the fit, and taking the largest difference from the nominal result as the uncertainty.
  We also evaluate a systematic uncertainty related to the assumption that $\kappa$ is independent of the proper decay length, by parametrizing $\kappa$ as a function of \ct using simulated samples. 
  A systematic uncertainty is assigned with the same methodology used to evaluate the ``model assumption'' systematic uncertainties, using the $\kappa(\ct)$ parametrization as an alternative hypothesis.

  \textit{Data/simulation difference:}
  The efficiency parametrization is found to be very sensitive to the muon and kaon \pt, and \PBzs meson rapidity distributions, hence a systematic uncertainty is assigned to cover the differences in each of these variables, between data and simulation. The effect is evaluated by reweighting the simulated distributions in each variable to agree with the data.
  The same weights are applied to the simulated samples used to estimate the efficiencies, which are then recomputed.
  The fit is repeated in each case and the sum in quadrature of the differences from the nominal result is taken as the systematic uncertainty.

  \textit{Flavor tagging:}
  The uncertainties associated with the flavor tagging are propagated by varying the parameters of the mistag probability calibration curves within their statistical uncertainties.
  For each variation, new calibration curves are produced and the data are refitted.
  The RMS of each fitted parameter distribution is then taken as the corresponding systematic uncertainty.
  We also evaluate the effect of the assumption that the signal and calibration channels have the same mistag calibration.
  The difference between the \PBzs and \PBpm calibrations is evaluated using simulated samples and is taken as the systematic uncertainty.
  The effect of the calibration function shape is evaluated by repeating the fit using a third-order polynomial and taking the difference with respect to the nominal result as the systematic uncertainty.
  The combined contribution of the three sources to the total systematic uncertainty is negligible.

  \textit{Different \wevt distribution in signal and background:}
  A systematic uncertainty is assigned to the possible differences in the mistag probabilities between signal and background.
  The separate signal and background \wevt distributions in data are first measured by using the \PBzs candidate invariant mass signal and sidebands regions.
  These distributions are separately modeled using the Kernel Density Estimation method~\cite{ref:rosenblatt1956,ref:parzen1962} and added to the fitting model.
  One thousand pseudo-experiments are generated and pull distributions with respect to the input values are used to evaluate the systematic uncertainty, as described in the ``model bias'' paragraph.

  \textit{Peaking background:}
  The systematic uncertainty related to the fixed yield of the peaking background component is evaluated by repeating the fit using a different yield obtained from a $\PBz \to \PJGy\, \kstar$ control sample in data.
  The difference with respect to the nominal result is taken as the systematic uncertainty.
  A systematic uncertainty is also assigned to the proper decay length modeling of the peaking background by forcing the lifetime to match the world-average value~\cite{ref:pdgTanabashi18prd}, repeating the fit, and taking the difference from the nominal result as the systematic uncertainty.

  \textit{$S$-$P$ wave interference:}
  The fit model does not take into account the difference in the invariant mass dependence between the $P$-wave from the \BsJpsiphi decay and the $S$-wave, which modifies their interference by a factor \ksp. 
  The corresponding systematic uncertainty is estimated using pseudo-experiments.
  The $k_\text{SP}$ factor is computed by integrating the $P$- and $S$-wave interference term in the \PGf candidate mass range, assuming that the $P$-wave amplitude is described by a relativistic Breit--Wigner distribution and the $S$-wave amplitude by a constant, and found to be $\ksp=0.54$.
  Different $S$-wave lineshapes are found to lead to very similar values of \ksp, with a variation no larger than ${\approx}$2\%.
  One thousand pseudo-experiments are generated applying $\ksp=0.54$ to the $i=8,\,9,\,10$ terms in Table~\ref{tab:kinematics} related to the $S$- and $P$-wave interference. 
  Pull distributions with respect to the input values are used to evaluate the systematic uncertainty, as described in the ``model bias'' paragraph.
  The parameters $\sqabs{\Aswav}$ and \Dms are the only ones whose total uncertainty is affected significantly by this approximation.

  \textit{$P(\sct)$ uncertainty:}
  In the fit to data the proper decay length uncertainty pdf is fixed to the one obtained from a pre-fit, as described in Section~\ref{sec:mlhfit}. 
  A systematic uncertainty is assigned by sampling this distribution 1000 times, using the parameter uncertainties obtained from the pre-fit. Each time the fit to data is repeated and the standard deviation of the obtained physics parameter distributions is taken as the systematic uncertainty.

A summary of the systematic uncertainties is given in Table~\ref{tab:systematics}. After adding the systematic uncertainties in quadrature, we measure the following values of the \CP-violating phase and the width difference between the two \PBzs mass eigenstates:
\begin{linenomath}
\begin{gather*}
  \phis = -11\pm 50\stat\pm 10\syst\mrad,\\
  \DGs = 0.114 \pm 0.014\stat\pm 0.007\syst\ips.
\end{gather*}
\end{linenomath}
{\tolerance=1000 The $\abs{\lambda}$ parameter is measured to be $\abs{\lambda}=0.972\pm 0.026\stat \pm 0.008\syst$, consistent with no direct \CP violation ($\abs{\lambda} = 1$).
The average of the heavy and light \PBzs mass eigenstate decay widths is determined to be $\Gs = 0.6531 \pm 0.0042\stat \pm 0.0026\syst\ips$, 
consistent with the world-average value $\Gs = 0.6624 \pm 0.0018\ips$~\cite{ref:pdgTanabashi18prd}.
The mass difference between the heavy and light \PBzs meson mass eigenstates is measured to be 
$\Dms = 17.51\,^{+\,0.10}_{-\,0.09}\stat\pm 0.03\syst\,\hps$, consistent with the theoretical prediction $\Dms = 18.77 \pm 0.86 \,\hps$~\cite{ref:Lenz2019lvd}, and in slight tension with the world-average value $\Dms = 17.757 \pm 0.021\,\hps$~\cite{ref:pdgTanabashi18prd}.
The uncertainties in all these measured parameters are dominated by the statistical component.
This analysis represents the first measurement by CMS of the mass difference $\Dms$ between the heavy and light \PBzs mass eigenstates and of the direct \CP observable $\abs{\lambda}$. \par}

\section{Combination with 8\texorpdfstring{\TeV}{ TeV} results}
\label{sec:combo}

The results presented in this Letter are in agreement with the earlier CMS result at a center-of-mass energy of 8\TeV~\cite{ref:cmsKhachatryan16plb}.
As explained in Section~\ref{sec:introd}, both measurements are performed with a similar number of events, with the one at $\sqrt{s} = 13 \TeV$ having a higher tagging efficiency.
This leads to an improvement in the uncertainty in quantities that require tagging, such as \phis, while but the uncertainties in those that do not use tagging, such as \DGs, depend on the raw number of events and are not improved relative to the $8\TeV$ result.
The two sets of results are combined using the BLUE method~\cite{ref:Lyons1988rp,ref:Valassi2003mu} as implemented in the \textsc{root} package~\cite{ref:Brun1997pa,ref:Nisius2014wua,ref:Nisius2020jmf} using the following physics parameters: \phis, \DGs, \Gs, $\sqabs{\Azero}$, $\sqabs{\Aperp}$, $\sqabs{\Aswav}$, \dpara, \dperp, and \dswpd.
The statistical correlations between the parameters obtained in each measurement are taken into account as well as the correlations of the systematic uncertainties discussed in Section~\ref{sec:result}.
Different sources of systematic uncertainties are assumed to be uncorrelated.
The systematic uncertainty correlation between the parameters of the $8\TeV$ result is assumed to be zero.
This assumption has been found to not impact the results in a noticeable way.
Since the muon tagging, the efficiency evaluation, and part of the fit model are different in the two measurements, the respective systematic uncertainties are treated as uncorrelated between the two sets of results.
The combined results for the \CP-violating phase and lifetime difference between the two mass eigenstates are:
\begin{linenomath}
\begin{gather*}
  \phis = \phisComb,\\
  \DGs = \DGsComb,
\end{gather*}
\end{linenomath}
with a correlation between the two parameters of $+0.02$.
The full combination results and the correlations between the various extracted parameters are reported in \suppMaterial.

\begin{figure}[htb]
  \centering
  \includegraphics[width=0.5\textwidth]{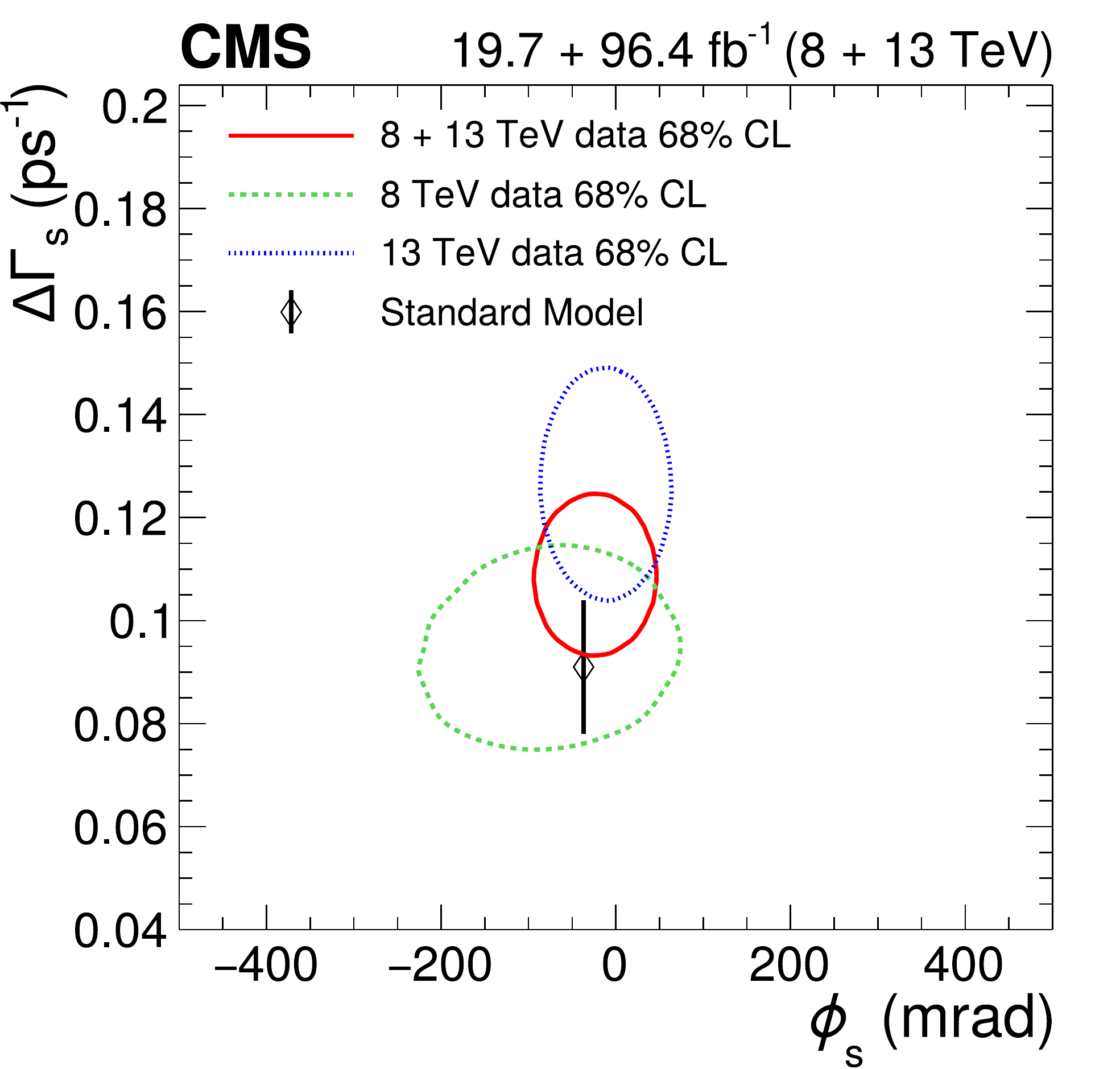}
  \caption{The two-dimensional likelihood contours at 68\% \CL\ in the $\phis$-$\DGs$ plane, for the CMS 8\TeV (dashed line), 13\TeV (dotted line), and combined (solid line) results.
  The contours for the individual results are obtained with likelihood scans, which are used to obtain the combined contour.
  In all contours only statistical uncertainties are taken into account.
  The SM prediction is shown with the diamond marker~\cite{ref:Charles11prd,ref:Lenz2019lvd}.}
  \label{fig:moneyplot}
\end{figure}

The two-dimensional $\phis$ vs. $\DGs$ likelihood contours at 68\% confidence level (\CL) for the individual and combined results, as well as the SM prediction, are shown in Fig.~\ref{fig:moneyplot}.
The contours for the individual results are obtained with likelihood scans, which are used to obtain the combined contour. The contours only account for the statistical uncertainty and the correlation between the two scanned variables, while the results from the combination obtained using the BLUE method take into account the statistical and systematic correlations of a wider range of variables.
The results are in agreement with each other and with the SM predictions.

\section{Summary}
\label{sec:conclu}

The \CP-violating phase $\phis$ and the decay width difference \DGs between the light and heavy \PBzs meson mass eigenstates are measured using a total of \Nev \BsJpsiphiMMKKalt signal events, collected by the CMS experiment at the LHC in proton-proton collisions at $\sqrt{s}=13\TeV$, corresponding to an integrated luminosity of \intL.
Events are selected using a trigger that requires an additional muon, which can be exploited to infer the flavor of the \PBzs meson at the time of production.
A novel opposite-side muon tagger based on deep neural networks has been developed to maximize the sensitivity of the present analysis. A high tagging power of ${\approx}$10\% is achieved, aided by the requirement of an additional muon in the signal sample imposed at the trigger level.

{\tolerance=900 The \CP-violating phase is measured to be $\phis = -11\pm 50\stat\pm 10\syst\mrad$, consistent both with the SM prediction $\phis = \phisSM \mrad$~\cite{ref:Charles11prd} and with the absence of \CP violation in the mixing-decay interference.
The decay width difference between the \PBzs mass eigenstates is measured to be $\DGs= 0.114 \pm 0.014\stat\pm 0.007\syst\ips$, consistent with the theoretical prediction $\DGs = 0.091 \pm 0.013\ips$~\cite{ref:Lenz2019lvd}.
In addition, the \CP-violating parameter $\abs{\lambda}$ and the average lifetime of the heavy and light \PBzs mass eigenstates, as well as their mass difference, have been measured.
The uncertainties in all these measurements are dominated by the statistical components. \par

The results presented in this Letter are further combined with those obtained by CMS at $\sqrt{s} = 8\TeV$~\cite{ref:cmsKhachatryan16plb}, yielding $\phis = \phisComb$ and $\DGs = \DGsComb$.
These results are significantly more precise than those from the previous CMS measurement at 8\TeV, and can be used to further constrain possible new-physics effects in \PBzs meson decay and mixing.}

\begin{acknowledgments}
  We congratulate our colleagues in the CERN accelerator departments for the excellent performance of the LHC and thank the technical and administrative staffs at CERN and at other CMS institutes for their contributions to the success of the CMS effort. In addition, we gratefully acknowledge the computing centers and personnel of the Worldwide LHC Computing Grid for delivering so effectively the computing infrastructure essential to our analyses. Finally, we acknowledge the enduring support for the construction and operation of the LHC and the CMS detector provided by the following funding agencies: BMBWF and FWF (Austria); FNRS and FWO (Belgium); CNPq, CAPES, FAPERJ, FAPERGS, and FAPESP (Brazil); MES (Bulgaria); CERN; CAS, MoST, and NSFC (China); COLCIENCIAS (Colombia); MSES and CSF (Croatia); RIF (Cyprus); SENESCYT (Ecuador); MoER, ERC IUT, PUT and ERDF (Estonia); Academy of Finland, MEC, and HIP (Finland); CEA and CNRS/IN2P3 (France); BMBF, DFG, and HGF (Germany); GSRT (Greece); NKFIA (Hungary); DAE and DST (India); IPM (Iran); SFI (Ireland); INFN (Italy); MSIP and NRF (Republic of Korea); MES (Latvia); LAS (Lithuania); MOE and UM (Malaysia); BUAP, CINVESTAV, CONACYT, LNS, SEP, and UASLP-FAI (Mexico); MOS (Montenegro); MBIE (New Zealand); PAEC (Pakistan); MSHE and NSC (Poland); FCT (Portugal); JINR (Dubna); MON, RosAtom, RAS, RFBR, and NRC KI (Russia); MESTD (Serbia); SEIDI, CPAN, PCTI, and FEDER (Spain); MOSTR (Sri Lanka); Swiss Funding Agencies (Switzerland); MST (Taipei); ThEPCenter, IPST, STAR, and NSTDA (Thailand); TUBITAK and TAEK (Turkey); NASU (Ukraine); STFC (United Kingdom); DOE and NSF (USA).
 
  \hyphenation{Rachada-pisek} Individuals have received support from the Marie-Curie program and the European Research Council and Horizon 2020 Grant, contract Nos.\ 675440, 752730, and 765710 (European Union); the Leventis Foundation; the A.P.\ Sloan Foundation; the Alexander von Humboldt Foundation; the Belgian Federal Science Policy Office; the Fonds pour la Formation \`a la Recherche dans l'Industrie et dans l'Agriculture (FRIA-Belgium); the Agentschap voor Innovatie door Wetenschap en Technologie (IWT-Belgium); the F.R.S.-FNRS and FWO (Belgium) under the ``Excellence of Science -- EOS" -- be.h project n.\ 30820817; the Beijing Municipal Science \& Technology Commission, No. Z191100007219010; the Ministry of Education, Youth and Sports (MEYS) of the Czech Republic; the Deutsche Forschungsgemeinschaft (DFG) under Germany's Excellence Strategy -- EXC 2121 ``Quantum Universe" -- 390833306; the Lend\"ulet (``Momentum") Program and the J\'anos Bolyai Research Scholarship of the Hungarian Academy of Sciences, the New National Excellence Program \'UNKP, the NKFIA research grants 123842, 123959, 124845, 124850, 125105, 128713, 128786, and 129058 (Hungary); the Council of Science and Industrial Research, India; the HOMING PLUS program of the Foundation for Polish Science, cofinanced from European Union, Regional Development Fund, the Mobility Plus program of the Ministry of Science and Higher Education, the National Science Center (Poland), contracts Harmonia 2014/14/M/ST2/00428, Opus 2014/13/B/ST2/02543, 2014/15/B/ST2/03998, and 2015/19/B/ST2/02861, Sonata-bis 2012/07/E/ST2/01406; the National Priorities Research Program by Qatar National Research Fund; the Ministry of Science and Higher Education, project no. 02.a03.21.0005 (Russia); the Programa Estatal de Fomento de la Investigaci{\'o}n Cient{\'i}fica y T{\'e}cnica de Excelencia Mar\'{\i}a de Maeztu, grant MDM-2015-0509 and the Programa Severo Ochoa del Principado de Asturias; the Thalis and Aristeia programs cofinanced by EU-ESF and the Greek NSRF; the Rachadapisek Sompot Fund for Postdoctoral Fellowship, Chulalongkorn University and the Chulalongkorn Academic into Its 2nd Century Project Advancement Project (Thailand); the Kavli Foundation; the Nvidia Corporation; the SuperMicro Corporation; the Welch Foundation, contract C-1845; and the Weston Havens Foundation (USA).
\end{acknowledgments}

\bibliography{auto_generated}

\providecommand{\href}[2]{#2}\begingroup\raggedright\begin{thebibliography}{10}%
\makeatletter
\providecommand{\hrefCMSnoop }[0]{\@secondoftwo}%
\makeatother
\providecommand{\doi}{\texttt{doi:}\begingroup \urlstyle{tt}\Url}

\bibitem{ref:Charles11prd}
\hrefCMSnoop {}{{The CKMfitter Group} Collaboration, ``{Predictions of selected
  flavour observables within the Standard Model}'',} \textit{ Phys. Rev. D}
  \textbf{ 84} (2011) 033005,
  \href{http://dx.doi.org/10.1103/PhysRevD.84.033005}{\doi{10.1103/PhysRevD.84.033005}},
  \href{http://www.arXiv.org/abs/1106.4041}{\texttt{arXiv:1106.4041}}. Updated
  with Summer 2019 results:
  \url{http://ckmfitter.in2p3.fr/www/results/plots\_summer19/num/ckmEval\_results\_summer19.pdf}.

\bibitem{Chiang:2009ev}
C.-W. Chiang\hrefCMSnoop {}{ {et~al.}, ``{New physics in $\PBzs \to \PJGy\PGf$:
  a general analysis}'',} \textit{ JHEP} \textbf{ 04} (2010) 031,
  \href{http://dx.doi.org/10.1007/JHEP04(2010)031}{\doi{10.1007/JHEP04(2010)031}},
  \href{http://www.arXiv.org/abs/0910.2929}{\texttt{arXiv:0910.2929}}.

\bibitem{ref:Artuso2015swg}
\hrefCMSnoop {}{M.~Artuso, G.~Borissov, and A.~Lenz, ``{CP violation in the
  $\PBzs$ system}'',} \textit{ Rev. Mod. Phys.} \textbf{ 88} (2016), no.~4,
  045002,
  \href{http://dx.doi.org/10.1103/RevModPhys.88.045002}{\doi{10.1103/RevModPhys.88.045002}},
  \href{http://www.arXiv.org/abs/1511.09466}{\texttt{arXiv:1511.09466}}.
  [Addendum: Rev.Mod.Phys. 91, 049901 (2019)].

\bibitem{ref:Lenz2019lvd}
\hrefCMSnoop {}{A.~Lenz and G.~Tetlalmatzi-Xolocotzi, ``{Model-independent
  bounds on new physics effects in non-leptonic tree-level decays of
  B-mesons}'',} \textit{ JHEP} \textbf{ 07} (2020) 177,
  \href{http://dx.doi.org/10.1007/JHEP07(2020)177}{\doi{10.1007/JHEP07(2020)177}},
  \href{http://www.arXiv.org/abs/1912.07621}{\texttt{arXiv:1912.07621}}.

\bibitem{ref:d0Abazov08prl}
\hrefCMSnoop {}{{D0} Collaboration, ``{Measurement of \PBzs mixing parameters
  from the flavor-tagged decay $\PBzs \to \PJGy\PGf$}'',} \textit{ Phys. Rev.
  Lett.} \textbf{ 101} (2008) 241801,
  \href{http://dx.doi.org/10.1103/PhysRevLett.101.241801}{\doi{10.1103/PhysRevLett.101.241801}},
\href{http://www.arXiv.org/abs/0802.2255}{\texttt{arXiv:0802.2255}}.

\bibitem{ref:d0Abazov12prd}
\hrefCMSnoop {}{{D0} Collaboration, ``{Measurement of the CP-violating phase
  $\phi_{\mathrm{s}}^{\PJGy\PGf}$ using the flavor-tagged decay $\PBzs \to
  \PJGy\PGf$ in 8\fbinv of p$\overline{\textrm{p}}$ collisions}'',} \textit{
  Phys. Rev. D} \textbf{ 85} (2012) 032006,
  \href{http://dx.doi.org/10.1103/PhysRevD.85.032006}{\doi{10.1103/PhysRevD.85.032006}},
\href{http://www.arXiv.org/abs/1109.3166}{\texttt{arXiv:1109.3166}}.

\bibitem{ref:cdfAaltonen08prl}
\hrefCMSnoop {}{{CDF} Collaboration, ``{First flavor-tagged determination of
  bounds on mixing-induced CP violation in $\PBzs \to \PJGy\PGf$ decays}'',}
  \textit{ Phys. Rev. Lett.} \textbf{ 100} (2008) 161802,
  \href{http://dx.doi.org/10.1103/PhysRevLett.100.161802}{\doi{10.1103/PhysRevLett.100.161802}},
\href{http://www.arXiv.org/abs/0712.2397}{\texttt{arXiv:0712.2397}}.

\bibitem{ref:cdfAaltonen12prd}
\hrefCMSnoop {}{{CDF} Collaboration, ``{Measurement of the CP-violating phase
  $\beta_{\mathrm{s}}^{\PJGy\PGf}$ in $\PBzs \to \PJGy\PGf$ decays with the CDF
  II detector}'',} \textit{ Phys. Rev. D} \textbf{ 85} (2012) 072002,
  \href{http://dx.doi.org/10.1103/PhysRevD.85.072002}{\doi{10.1103/PhysRevD.85.072002}},
\href{http://www.arXiv.org/abs/1112.1726}{\texttt{arXiv:1112.1726}}.

\bibitem{ref:cdfAaltonen12prl}
\hrefCMSnoop {}{{CDF} Collaboration, ``{Measurement of the bottom-strange meson
  mixing phase in the full CDF data set}'',} \textit{ Phys. Rev. Lett.}
  \textbf{ 109} (2012) 171802,
  \href{http://dx.doi.org/10.1103/PhysRevLett.109.171802}{\doi{10.1103/PhysRevLett.109.171802}},
\href{http://www.arXiv.org/abs/1208.2967}{\texttt{arXiv:1208.2967}}.

\bibitem{ref:atlasAad12jhep}
\hrefCMSnoop {}{{ATLAS Collaboration}, ``{Time-dependent angular analysis of
  the decay $\PBzs \to \PJGy\PGf$ and extraction of $\DGs$ and the CP-violating
  weak phase $\phis$ by ATLAS}'',} \textit{ JHEP} \textbf{ 12} (2012) 072,
  \href{http://dx.doi.org/10.1007/JHEP12(2012)072}{\doi{10.1007/JHEP12(2012)072}},
\href{http://www.arXiv.org/abs/1208.0572}{\texttt{arXiv:1208.0572}}.

\bibitem{ref:atlasAad14prd}
\hrefCMSnoop {}{{ATLAS Collaboration}, ``{Flavor tagged time-dependent angular
  analysis of the $\PBzs \to \PJGy\PGf$ decay and extraction of $\DGs$ and the
  weak phase $\phis$ in ATLAS}'',} \textit{ Phys. Rev. D} \textbf{ 90} (2014)
  052007,
  \href{http://dx.doi.org/10.1103/PhysRevD.90.052007}{\doi{10.1103/PhysRevD.90.052007}},
\href{http://www.arXiv.org/abs/1407.1796}{\texttt{arXiv:1407.1796}}.

\bibitem{ref:atlasAad16jhep}
\hrefCMSnoop {}{{ATLAS Collaboration}, ``{Measurement of the CP-violating phase
  $\phis$ and the \PBs meson decay width difference with $\PBzs \to \PJGy\PGf$
  decays in ATLAS}'',} \textit{ JHEP} \textbf{ 12} (2016) 072,
  \href{http://dx.doi.org/10.1007/JHEP08(2016)147}{\doi{10.1007/JHEP08(2016)147}},
\href{http://www.arXiv.org/abs/1601.03297}{\texttt{arXiv:1601.03297}}.

\bibitem{ref:Aad:2020jfw}
\hrefCMSnoop {}{{ATLAS Collaboration}, ``{Measurement of the $CP$-violating
  phase $\phi_s$ in $\PBzs \to \PJGy\PGf$ decays in ATLAS at 13 TeV}'',}
  (2020).
  \href{http://www.arXiv.org/abs/2001.07115}{\texttt{arXiv:2001.07115}}.
Submitted to EPJC.

\bibitem{ref:cmsKhachatryan16plb}
\hrefCMSnoop {}{{CMS Collaboration}, ``{Measurement of the CP-violating weak
  phase \phis\ and the decay width difference \DGs\ using the \BsJpsiphi \
  decay channel in $\Pp\Pp$ collisions at $\sqrt{s} = 8$ TeV}'',} \textit{
  Phys. Lett. B} \textbf{ 757} (2016) 97,
  \href{http://dx.doi.org/10.1016/j.physletb.2016.03.046}{\doi{10.1016/j.physletb.2016.03.046}},
\href{http://www.arXiv.org/abs/1507.07527}{\texttt{arXiv:1507.07527}}.

\bibitem{ref:lhcbAaij12plb}
\hrefCMSnoop {}{{LHCb Collaboration}, ``{Measurement of the CP-violating phase
  $\phis$ in $\PABzs \to \PJGy\pi^+\pi^-$ decays}'',} \textit{ Phys. Lett. B}
  \textbf{ 713} (2012) 378,
  \href{http://dx.doi.org/10.1016/j.physletb.2012.06.032}{\doi{10.1016/j.physletb.2012.06.032}},
\href{http://www.arXiv.org/abs/1204.5675}{\texttt{arXiv:1204.5675}}.

\bibitem{ref:lhcbAaij13prd}
\hrefCMSnoop {}{{LHCb Collaboration}, ``{Measurement of CP violation and the
  $\PBzs $ meson decay width difference with $\PBzs \to
  \PJGy\mathrm{K}^+\mathrm{K}^-$ and $\PBzs \to \PJGy\pi^+\pi^-$ decays}'',}
  \textit{ Phys. Rev. D} \textbf{ 87} (2013) 112010,
  \href{http://dx.doi.org/10.1103/PhysRevD.87.112010}{\doi{10.1103/PhysRevD.87.112010}},
\href{http://www.arXiv.org/abs/1304.2600}{\texttt{arXiv:1304.2600}}.

\bibitem{ref:lhcbAaij14plb}
\hrefCMSnoop {}{{LHCb Collaboration}, ``{Measurement of the CP-violating phase
  $\phis$ in $ \PABzs\to \PJGy\pi^+\pi^-$ decays}'',} \textit{ Phys. Lett. B}
  \textbf{ 736} (2014) 186,
  \href{http://dx.doi.org/10.1016/j.physletb.2014.06.079}{\doi{10.1016/j.physletb.2014.06.079}},
\href{http://www.arXiv.org/abs/1405.4140}{\texttt{arXiv:1405.4140}}.

\bibitem{ref:lhcbAaij15prl}
\hrefCMSnoop {}{{LHCb Collaboration}, ``{Precision measurement of CP violation
  in $\PBzs \to \PJGy\mathrm{K}^+\mathrm{K}^-$ decays}'',} \textit{ Phys. Rev.
  Lett.} \textbf{ 114} (2015) 041801,
  \href{http://dx.doi.org/10.1103/PhysRevLett.114.041801}{\doi{10.1103/PhysRevLett.114.041801}},
\href{http://www.arXiv.org/abs/1411.3104}{\texttt{arXiv:1411.3104}}.

\bibitem{ref:Aaij2019vot}
\hrefCMSnoop {}{{LHCb Collaboration}, ``{Updated measurement of time-dependent
  {$CP$}-violating observables in $B^{0}_{s}\to J/\psi K^+ K^-$ decays}'',}
  \textit{ Eur. Phys. J. C} \textbf{ 79} (2019), no.~8, 706,
  \href{http://dx.doi.org/10.1140/epjc/s10052-019-7159-8}{\doi{10.1140/epjc/s10052-019-7159-8}},
  \href{http://www.arXiv.org/abs/1906.08356}{\texttt{arXiv:1906.08356}}.
  [Erratum: \DOI{10.1140/epjc/s10052-020-7875-0}].

\bibitem{ref:lhcbAaij16plb}
\hrefCMSnoop {}{{LHCb Collaboration}, ``{First study of the CP-violating phase
  and decay-width difference in $\PBzs \to \Pgy\PGf$}'',} \textit{ Phys. Lett.
  B} \textbf{ 762} (2016) 253,
  \href{http://dx.doi.org/10.1016/j.physletb.2016.09.028}{\doi{10.1016/j.physletb.2016.09.028}},
\href{http://www.arXiv.org/abs/1608.04855}{\texttt{arXiv:1608.04855}}.

\bibitem{ref:lhcbAaij14prl}
\hrefCMSnoop {}{{LHCb Collaboration}, ``{Measurement of the CP-violating phase
  $\phis$ in $\PABzs \to \PsDp\PsDm$ decays}'',} \textit{ Phys. Rev. Lett.}
  \textbf{ 113} (2014) 211801,
  \href{http://dx.doi.org/10.1103/PhysRevLett.113.211801}{\doi{10.1103/PhysRevLett.113.211801}},
\href{http://www.arXiv.org/abs/1409.4619}{\texttt{arXiv:1409.4619}}.

\bibitem{ref:Dighe99epjc}
\hrefCMSnoop {}{A.~S. Dighe, I.~Dunietz, and R.~Fleischer, ``{Extracting CKM
  phases and $\PBzs$--$\PABzs$ mixing parameters from angular distributions of
  non-leptonic B decays}'',} \textit{ Eur. Phys. J. C} \textbf{ 6} (1999) 647,
  \href{http://dx.doi.org/10.1007/s100529800954}{\doi{10.1007/s100529800954}},
  \href{http://www.arXiv.org/abs/hep-ph/9804253}{\texttt{arXiv:hep-ph/9804253}}.

\bibitem{ref:Dighe95plb}
\hrefCMSnoop {}{A.~S. Dighe, I.~Dunietz, H.~J. Lipkin, and J.~L. Rosner,
  ``{Angular distributions and lifetime differences in $\PBzs \to \PJGy\PGf$
  decays}'',} \textit{ Phys. Lett. B} \textbf{ 369} (1996) 144,
  \href{http://dx.doi.org/10.1016/0370-2693(95)01523-X}{\doi{10.1016/0370-2693(95)01523-X}},
\href{http://www.arXiv.org/abs/hep-ph/9511363}{\texttt{arXiv:hep-ph/9511363}}.

\bibitem{ref:Branco99irmp}
G.~C. Branco, L.~Lavoura, and J.~P. Silva, ``{CP Violation}'', volume 103 of
  \textit{ International Series of Monographs on Physics}.
\newblock Clarendon Press, Oxford, UK, 1999.
\newblock ISBN~0198503997.

\bibitem{ref:cmsmpr18}
\hrefCMSnoop {}{{CMS Collaboration}, ``{Performance of the CMS muon detector
  and muon reconstruction with proton-proton collisions at $\sqrt{s} = 13$
  TeV}'',} \textit{ JINST} \textbf{ 13} (2018) P06015,
  \href{http://dx.doi.org/10.1088/1748-0221/13/06/P06015}{\doi{10.1088/1748-0221/13/06/P06015}},
  \href{http://www.arXiv.org/abs/1804.04528}{\texttt{arXiv:1804.04528}}.

\bibitem{ref:cmstri17}
\hrefCMSnoop {}{{CMS Collaboration}, ``{The CMS trigger system}'',} \textit{
  JINST} \textbf{ 12} (2017) P01020,
  \href{http://dx.doi.org/10.1088/1748-0221/12/01/P01020}{\doi{10.1088/1748-0221/12/01/P01020}},
  \href{http://www.arXiv.org/abs/1609.02366}{\texttt{arXiv:1609.02366}}.

\bibitem{ref:cmsexp08}
\hrefCMSnoop {}{{CMS Collaboration}, ``{The {CMS} Experiment at the {CERN}
  {LHC}}'',} \textit{ JINST} \textbf{ 3} (2008) S08004,
  \href{http://dx.doi.org/10.1088/1748-0221/3/08/S08004}{\doi{10.1088/1748-0221/3/08/S08004}}.

\bibitem{ref:Antcheva09cpc}
\hrefCMSnoop {}{I.~Antcheva {et~al.}, ``{ROOT --- A C++ framework for petabyte
  data storage, statistical analysis and visualization}'',} \textit{ Comput.
  Phys. Commun.} \textbf{ 180} (2009) 2499,
  \href{http://dx.doi.org/10.1016/j.cpc.2009.08.005}{\doi{10.1016/j.cpc.2009.08.005}},
  \href{http://www.arXiv.org/abs/1508.07749}{\texttt{arXiv:1508.07749}}.

\bibitem{ref:tmvaHocker07pos}
\hrefCMSnoop {}{H.~Voss, A.~H{\"o}cker, J.~Stelzer, and F.~Tegenfeldt,
  ``{TMVA}, the toolkit for multivariate data analysis with {ROOT}'',} in
  \textit{ XIth International Workshop on Advanced Computing and Analysis
  Techniques in Physics Research (ACAT)}, p.~40.
\newblock 2007.
\newblock
  \href{http://www.arXiv.org/abs/physics/0703039}{\texttt{arXiv:physics/0703039}}.
\newblock {[PoS(ACAT)040]}.
\href{http://dx.doi.org/10.22323/1.050.0040}{\doi{10.22323/1.050.0040}}.

\bibitem{ref:Fruhwirth87nim}
\hrefCMSnoop {}{R.~Fr{\"u}hwirth, ``{Application of Kalman filtering to track
  and vertex fitting}'',} \textit{ Nucl. Instrum. Meth. A} \textbf{ 262} (1987)
  444,
\href{http://dx.doi.org/10.1016/0168-9002(87)90887-4}{\doi{10.1016/0168-9002(87)90887-4}}.

\bibitem{ref:pdgTanabashi18prd}
\hrefCMSnoop {}{{Particle Data Group}, M.~Tanabashi {et~al.}, ``Review of
  particle physics'',} \textit{ Phys. Rev. D} \textbf{ 98} (2018) 030001,
  \href{http://dx.doi.org/10.1103/PhysRevD.98.030001}{\doi{10.1103/PhysRevD.98.030001}}.

\bibitem{Chatrchyan:2014fea}
\hrefCMSnoop {}{{CMS Collaboration}, ``{Description and performance of track
  and primary-vertex reconstruction with the {CMS} tracker}'',} \textit{ JINST}
  \textbf{ 9} (2014) P10009,
  \href{http://dx.doi.org/10.1088/1748-0221/9/10/P10009}{\doi{10.1088/1748-0221/9/10/P10009}},
\href{http://www.arXiv.org/abs/1405.6569}{\texttt{arXiv:1405.6569}}.

\bibitem{Sjostrand:2014zea}
T.~Sj{\"o}strand\hrefCMSnoop {}{ {et~al.}, ``{An introduction to PYTHIA
  8.2}'',} \textit{ Comput. Phys. Commun.} \textbf{ 191} (2015) 159,
  \href{http://dx.doi.org/10.1016/j.cpc.2015.01.024}{\doi{10.1016/j.cpc.2015.01.024}},
\href{http://www.arXiv.org/abs/1410.3012}{\texttt{arXiv:1410.3012}}.

\bibitem{Sirunyan:2019dfx}
\hrefCMSnoop {}{{CMS Collaboration}, ``{Extraction and validation of a new set
  of CMS PYTHIA8 tunes from underlying-event measurements}'',} \textit{ Eur.
  Phys. J. C} \textbf{ 80} (2020) 4,
  \href{http://dx.doi.org/10.1140/epjc/s10052-019-7499-4}{\doi{10.1140/epjc/s10052-019-7499-4}},
  \href{http://www.arXiv.org/abs/1903.12179}{\texttt{arXiv:1903.12179}}.

\bibitem{Ball:2017nwa}
\hrefCMSnoop {}{{NNPDF} Collaboration, ``{Parton distributions from
  high-precision collider data}'',} \textit{ Eur. Phys. J. C} \textbf{ 77}
  (2017) 663,
  \href{http://dx.doi.org/10.1140/epjc/s10052-017-5199-5}{\doi{10.1140/epjc/s10052-017-5199-5}},
  \href{http://www.arXiv.org/abs/1706.00428}{\texttt{arXiv:1706.00428}}.

\bibitem{ref:Lange01nim}
\hrefCMSnoop {}{D.~J. Lange, ``{The EvtGen particle decay simulation
  package}'',} \textit{ Nucl. Instrum. Meth. A} \textbf{ 462} (2001) 152,
\href{http://dx.doi.org/10.1016/S0168-9002(01)00089-4}{\doi{10.1016/S0168-9002(01)00089-4}}.

\bibitem{ref:Barberio91cpc}
\hrefCMSnoop {}{E.~Barberio, B.~van Eijk, and Z.~W\c{a}s, ``{{\PHOTOS} --- a
  universal Monte Carlo for QED radiative corrections in decays}'',} \textit{
  Comput. Phys. Commun.} \textbf{ 66} (1991) 115,
  \href{http://dx.doi.org/10.1016/0010-4655(91)90012-A}{\doi{10.1016/0010-4655(91)90012-A}}.

\bibitem{ref:Barberio94cpc}
\hrefCMSnoop {}{E.~Barberio and Z.~W\c{a}s, ``{{\PHOTOS} --- a universal Monte
  Carlo for QED radiative corrections: version 2.0}'',} \textit{ Comput. Phys.
  Commun.} \textbf{ 79} (1994) 291,
  \href{http://dx.doi.org/10.1016/0010-4655(94)90074-4}{\doi{10.1016/0010-4655(94)90074-4}}.

\bibitem{ref:Agostinelli03nim}
\hrefCMSnoop {}{{GEANT4} Collaboration, ``{{\GEANTfour} --- a simulation
  toolkit}'',} \textit{ Nucl. Instrum. Meth. A} \textbf{ 506} (2003) 250,
\href{http://dx.doi.org/10.1016/S0168-9002(03)01368-8}{\doi{10.1016/S0168-9002(03)01368-8}}.

\bibitem{ref:chollet2015keras}
\hrefCMSnoop {}{F.~Chollet {et~al.}, ``Keras''.} \url{https://keras.io}, 2015.

\bibitem{Sirunyan_2017}
\hrefCMSnoop {}{{CMS Collaboration}, ``{Particle-flow reconstruction and global
  event description with the CMS detector}'',} \textit{ JINST} \textbf{ 12}
  (2017) P10003,
  \href{http://dx.doi.org/10.1088/1748-0221/12/10/P10003}{\doi{10.1088/1748-0221/12/10/P10003}},
\href{http://www.arXiv.org/abs/1706.04965}{\texttt{arXiv:1706.04965}}.

\bibitem{kingma2014adam}
\hrefCMSnoop {}{D.~Kingma and J.~Ba, ``Adam: A method for stochastic
  optimization'',} (12, 2014).
  \href{http://www.arXiv.org/abs/1412.6980}{\texttt{arXiv:1412.6980}}.

\bibitem{Verkerke:2003ir}
\hrefCMSnoop {}{W.~Verkerke and D.~Kirkby, ``The roofit toolkit for data
  modeling'',} 2003.

\bibitem{ref:Aaij2012eq}
\hrefCMSnoop {}{{LHCb Collaboration}, ``{Determination of the sign of the decay
  width difference in the $B_s$ system}'',} \textit{ Phys. Rev. Lett.} \textbf{
  108} (2012) 241801,
  \href{http://dx.doi.org/10.1103/PhysRevLett.108.241801}{\doi{10.1103/PhysRevLett.108.241801}},
  \href{http://www.arXiv.org/abs/1202.4717}{\texttt{arXiv:1202.4717}}.

\bibitem{Johnson:1949zj}
\hrefCMSnoop {}{N.~L. Johnson, ``{Systems of frequency curves generated by
  methods of translation}'',} \textit{ Biometrika} \textbf{ 36} (1949) 149,
\href{http://dx.doi.org/10.1093/biomet/36.1-2.149}{\doi{10.1093/biomet/36.1-2.149}}.

\bibitem{ref:rosenblatt1956}
\hrefCMSnoop {}{M.~Rosenblatt, ``Remarks on some nonparametric estimates of a
  density function'',} \textit{ Ann. Math. Statist.} \textbf{ 27} (1956) 832,
  \href{http://dx.doi.org/10.1214/aoms/1177728190}{\doi{10.1214/aoms/1177728190}}.

\bibitem{ref:parzen1962}
\hrefCMSnoop {}{E.~Parzen, ``On estimation of a probability density function
  and mode'',} \textit{ Ann. Math. Statist.} \textbf{ 33} (1962) 1065,
  \href{http://dx.doi.org/10.1214/aoms/1177704472}{\doi{10.1214/aoms/1177704472}}.

\bibitem{ref:Lyons1988rp}
\hrefCMSnoop {}{L.~Lyons, D.~Gibaut, and P.~Clifford, ``{How to combine
  correlated estimates of a single physical quantity}'',} \textit{ Nucl.
  Instrum. Meth. A} \textbf{ 270} (1988) 110,
  \href{http://dx.doi.org/10.1016/0168-9002(88)90018-6}{\doi{10.1016/0168-9002(88)90018-6}}.

\bibitem{ref:Valassi2003mu}
\hrefCMSnoop {}{A.~Valassi, ``{Combining correlated measurements of several
  different physical quantities}'',} \textit{ Nucl. Instrum. Meth. A} \textbf{
  500} (2003) 391,
  \href{http://dx.doi.org/10.1016/S0168-9002(03)00329-2}{\doi{10.1016/S0168-9002(03)00329-2}}.

\bibitem{ref:Brun1997pa}
\hrefCMSnoop {}{R.~Brun and F.~Rademakers, ``{ROOT: An object oriented data
  analysis framework}'',} \textit{ Nucl. Instrum. Meth. A} \textbf{ 389} (1997)
  81,
  \href{http://dx.doi.org/10.1016/S0168-9002(97)00048-X}{\doi{10.1016/S0168-9002(97)00048-X}}.

\bibitem{ref:Nisius2014wua}
\hrefCMSnoop {}{R.~Nisius, ``{On the combination of correlated estimates of a
  physics observable}'',} \textit{ Eur. Phys. J. C} \textbf{ 74} (2014), no.~8,
  3004,
  \href{http://dx.doi.org/10.1140/epjc/s10052-014-3004-2}{\doi{10.1140/epjc/s10052-014-3004-2}},
  \href{http://www.arXiv.org/abs/1402.4016}{\texttt{arXiv:1402.4016}}.

\bibitem{ref:Nisius2020jmf}
\hrefCMSnoop {}{R.~Nisius, ``{BLUE: combining correlated estimates of physics
  observables within ROOT using the Best Linear Unbiased Estimate method}'',}
  \textit{ SoftwareX} \textbf{ 11} (1, 2020) 100468,
  \href{http://dx.doi.org/10.1016/j.softx.2020.100468}{\doi{10.1016/j.softx.2020.100468}},
  \href{http://www.arXiv.org/abs/2001.10310}{\texttt{arXiv:2001.10310}}.

\end{thebibliography}\endgroup
\ifthenelse{\boolean{cms@external}}{}{
\clearpage
\appendix
\numberwithin{table}{section}
\numberwithin{figure}{section}
\section{Supplemental material: correlation matrices and additional results\label{app:suppMat}}

\label{sec:appendix}
\begin{table}[!ht]
  \centering
  \topcaption{Statistical correlation matrix between the physics parameters as obtained from the ML fit to the 13\TeV data.}
  \cmsTable{
  \begin{tabular}{c|ccccccccccc}
       & $\phis$ & $\DGs$ & $\Dms$ & $\abs{\lambda}$ & $\Gs$ & $\sqabs{\Azero}$ & $\sqabs{\Aperp}$ & $\sqabs{\Aswav}$ & $\dpara$ & $\dperp$ & $\dswpd$\\
      \hline
      $\phis$ & $+1.00$ & $-0.02$ & $-0.19$ & $+0.22$ & $\hphantom{+}0.00$ & $-0.01$ & $+0.01$ & $-0.01$ & $-0.02$ & $-0.09$ & $+0.03$\\
      $\DGs$ &  & $+1.00$ & $-0.02$ & $\hphantom{+}0.00$ & $-0.48$ & $+0.63$ & $-0.71$ & $\hphantom{+}0.00$ & $+0.01$ & $-0.01$ & $-0.04$\\
      $\Dms$ &  &  & $+1.00$ & $-0.14$ & $+0.03$ & $-0.01$ & $+0.02$ & $+0.03$ & $+0.01$ & $+0.68$ & $-0.05$\\
      $\abs{\lambda}$ &  &  &  & $+1.00$ & $-0.02$ & $\hphantom{+}0.00$ & $-0.01$ & $-0.03$ & $-0.06$ & $-0.18$ & $+0.05$\\
      $\Gs$ &  &  &  &  & $+1.00$ & $-0.31$ & $+0.42$ & $+0.15$ & $-0.02$ & $+0.02$ & $-0.05$\\
      $\sqabs{\Azero}$ &  &  &  &  &  & $+1.00$ & $-0.61$ & $+0.15$ & $-0.01$ & $-0.01$ & $-0.09$\\
      $\sqabs{\Aperp}$ &  &  &  &  &  &  & $+1.00$ & $-0.11$ & $-0.06$ & $\hphantom{+}0.00$ & $+0.06$\\
      $\sqabs{\Aswav}$ &  &  &  &  &  &  &  & $+1.00$ & $-0.07$ & $+0.02$ & $-0.44$\\
      $\dpara$ &  &  &  &  &  &  &  &  & $+1.00$ & $+0.27$ & $-0.02$\\
      $\dperp$ &  &  &  &  &  &  &  &  &  & $+1.00$ & $-0.10$\\
      $\dswpd$ &  &  &  &  &  &  &  &  &  &  & $+1.00$\\
  \end{tabular}
  }
  \label{tab:corr13}
\end{table}

\begin{table}[!ht]
  \centering
  \topcaption{Physics parameters values as obtained from the combination between the CMS 8\TeV and 13\TeV results.}
  \begin{tabular}{cr@{}lll}
    Parameter & \multicolumn{2}{c}{Value} & \multicolumn{1}{c}{Stat. uncer.} & \multicolumn{1}{c}{Syst. uncer.}\\
    \hline 
    $\phis$~[mrad]      & $-21$ &         & $\pm\, 44$      & $\pm\, 10$ \\
    $\DGs$~[ps$^{-1}$]  & $0$   & $.1032$ & $\pm\, 0.0095$  & $\pm\, 0.0048$ \\
    $\Gs$~[ps$^{-1}$]   & $0$   & $.6590$ & $\pm\, 0.0032$  & $\pm\, 0.0023$ \\
    $\sqabs{\Azero}$    & $0$   & $.5289$ & $\pm\, 0.0038$  & $\pm\, 0.0041$ \\
    $\sqabs{\Aperp}$    & $0$   & $.2393$ & $\pm\, 0.0050$  & $\pm\, 0.0037$ \\
    $\sqabs{\Aswav}$    & $0$   & $.016$  & $\pm\, 0.006$   & $\pm\, 0.013$ \\
    $\dpara$~[rad]      & $3$   & $.19$   & $\pm\, 0.12$    & $\pm\, 0.04$ \\
    $\dperp$~[rad]      & $2$   & $.78$   & $\pm\, 0.15$    & $\pm\, 0.06$ \\
    $\dswpd$~[rad]      & $0$   & $.238$  & $\pm\, 0.078$   & $\pm\, 0.046$ \\
  \end{tabular}
  \label{tab:results8_13}
\end{table}

\begin{table}[!th]
  \centering
  \topcaption{Correlations between the physics parameters as obtained from the combination between the CMS 8\TeV and 13\TeV results. Correlations are both statistical and systematic.}
  \begin{tabular}{c|ccccccccc}
     & $\phis$ & $\DGs$ & $\Gs$ & $\sqabs{\Azero}$ & $\sqabs{\Aperp}$ & $\sqabs{\Aswav}$ & $\dpara$ & $\dperp$ & $\dswpd$\\
    \hline
    $\phis$ & $+1.00$ & $+0.02$ & $-0.03$ & $+0.01$ & $-0.01$ & $+0.01$ & $-0.01$ & $-0.08$ & $+0.03$\\
    $\DGs$ &  & $+1.00$ & $-0.45$ & $+0.43$ & $-0.57$ & $+0.01$ & $+0.01$ & $\hphantom{+}0.00$ & $-0.01$\\
    $\Gs$ &  &  & $+1.00$ & $-0.17$ & $+0.30$ & $+0.06$ & $-0.03$ & $\hphantom{+}0.00$ & $-0.08$\\
    $\sqabs{\Azero}$ &  &  &  & $+1.00$ & $-0.56$ & $+0.25$ & $-0.03$ & $+0.01$ & $-0.18$\\
    $\sqabs{\Aperp}$ &  &  &  &  & $+1.00$ & $-0.08$ & $-0.03$ & $+0.01$ & $+0.14$\\
    $\sqabs{\Aswav}$ &  &  &  &  &  & $+1.00$ & $-0.02$ & $+0.02$ & $-0.20$\\
    $\dpara$ &  &  &  &  &  &  & $+1.00$ & $+0.26$ & $\hphantom{+}0.00$\\
    $\dperp$ &  &  &  &  &  &  &  & $+1.00$ & $-0.05$\\
    $\dswpd$ &  &  &  &  &  &  &  &  & $+1.00$\\
  \end{tabular}
  \label{tab:corr8_13}
\end{table}

}
\cleardoublepage \section{The CMS Collaboration \label{app:collab}}\begin{sloppypar}\hyphenpenalty=5000\widowpenalty=500\clubpenalty=5000\vskip\cmsinstskip
\textbf{Yerevan Physics Institute, Yerevan, Armenia}\\*[0pt]
A.M.~Sirunyan$^{\textrm{\dag}}$, A.~Tumasyan
\vskip\cmsinstskip
\textbf{Institut f\"{u}r Hochenergiephysik, Wien, Austria}\\*[0pt]
W.~Adam, F.~Ambrogi, T.~Bergauer, M.~Dragicevic, J.~Er\"{o}, A.~Escalante~Del~Valle, R.~Fr\"{u}hwirth\cmsAuthorMark{1}, M.~Jeitler\cmsAuthorMark{1}, N.~Krammer, L.~Lechner, D.~Liko, T.~Madlener, I.~Mikulec, F.M.~Pitters, N.~Rad, J.~Schieck\cmsAuthorMark{1}, R.~Sch\"{o}fbeck, M.~Spanring, S.~Templ, W.~Waltenberger, C.-E.~Wulz\cmsAuthorMark{1}, M.~Zarucki
% ...
\vskip\cmsinstskip
\dag: Deceased\\
1:  Also at Vienna University of Technology, Vienna, Austria\\
% ...
\end{sloppypar}
\end{document}